 \documentclass[onecolumn]{nst}
 
 \usepackage{amsthm,amsmath,amssymb}
 \usepackage{mathrsfs}
 
\usepackage{subfigure,dcolumn}
\usepackage{epstopdf}
\usepackage{mhchem}
\usepackage{upgreek}

 \makeatother
\begin{document}

\title{Recent developments in chiral and spin polarization effects in heavy-ion collisions}

\author{Jian-Hua Gao}
\email[Corresponding author, ]{gaojh@sdu.edu.cn}

\affiliation{Shandong Key Laboratory of Optical Astronomy and Solar-Terrestrial
Environment, School of Space Science and Physics, Institute of Space
Sciences, Shandong University, Weihai 264209, China}
\author{Guo-Liang Ma}
\email[Corresponding author, ]{glma@fudan.edu.cn}

\affiliation{Key Laboratory of Nuclear Physics and Ion-beam Application (MOE),
Institute of Modern Physics, Fudan University, Shanghai 200433, China}
\author{Shi Pu}
\email[Corresponding author, ]{shipu@ustc.edu.cn}

\affiliation{Department of Modern Physics, University of Science and Technology
of China, Hefei 230026, China}

\author{Qun Wang}
\email[Corresponding author, ]{qunwang@ustc.edu.cn}
\affiliation{Department of Modern Physics, University of Science and Technology
of China, Hefei 230026, China}

\begin{abstract}
We give a brief overview of recent theoretical and experimental results on the chiral magnetic effect and spin polarization
effect in heavy-ion collisions. We present updated experimental
results for the chiral magnetic effect and related phenomena. The
time evolution of the magnetic fields in different models is discussed.
The newly developed quantum kinetic theory for massive fermions is
reviewed. We present theoretical and experimental results
for the polarization of $\Lambda$ hyperons and the $\rho_{00}$ value of vector
mesons. 
\end{abstract}


\maketitle

\section{Introduction \label{sec:Introduction}}

In relativistic heavy-ion collisions, two charged nuclei collide to
produce a hot, dense state of matter known as a quark-gluon plasma (QGP).
Very high magnetic fields and orbital angular momenta (OAMs) are generated
in these collisions. The magnetic fields are on the order of $10^{17-18}$ Gs
\citep{Bloczynski:2012en,Deng:2012pc,Roy:2015coa,Li:2016tel} and are
the strongest magnetic fields observed in nature.
The QGP is also found to be the most vortical fluid \citep{STAR:2017ckg},
where a huge OAM is transferred to the fluid in the form of vorticity
fields. These novel phenomena open a new window for the study of the QGP
in heavy-ion collisions. These novel phenomena are quantum in nature
and are usually negligible in classical fluids. 

The chiral magnetic effect (CME) and chiral separation effect (CSE)
\citep{Kharzeev:2004ey,Kharzeev:2007jp,Fukushima:2008xe} are two
quantum effects on the magnetic field in a chiral fermion system. In the CME,
a charge current is induced along the magnetic field: 
\begin{eqnarray}
\boldsymbol{j} & = & \frac{e^{2}}{2\pi^{2}}\mu_{5}\boldsymbol{B},\label{eq:CME}
\end{eqnarray}
where $\mu_{5}$ is the chiral chemical potential. The magnetic field
can also generate a chiral current 
\begin{equation}
\boldsymbol{j}_{5}=\frac{e^{2}}{2\pi^{2}}\mu\boldsymbol{B},\label{eq:CSE}
\end{equation}
where $\mu$ is the charge chemical potential. These two quantum effects
are related to the chiral anomaly, which is absent from classical theories.
The chiral chemical potential represents a chirality imbalance. The
nonzero $\mu_{5}$ arises from topological fluctuations in quantum
chromodynamics (QCD), which are related to the local violation of parity 
and charge parity. Therefore, the observation of the CME
in heavy-ion collisions implies local parity and charge parity violation. The
collective modes associated with the CME and CSE are called the chiral
magnetic wave (CMW). In addition, the CME has been observed in condensed
matter \citep{Li:2014bha} and can be applied in quantum computing
\citep{Kharzeev:2019ceh}. These phenomena are discussed in more detail in recent
reviews \citep{Kharzeev:2015znc,Liao:2014ava,Miransky:2015ava,Huang:2015oca,Skokov:2016yrj,Fukushima:2018grm,Bzdak:2019pkr,Zhao:2019hta,Wang:2018ygc,Liu:2020ymh}
and references therein. 

In non-central heavy-ion collisions, part of the very large OAM of colliding
nuclei is converted to a vorticity field in the fluid. The vorticity
can be regarded as the local OAM of the fluid and can polarize particles
with spins through spin--orbit coupling. The global spin polarization
of particles is along the direction of the global OAM of colliding
nuclei (or the direction perpendicular to the reaction plane). The
global polarization of $\Lambda$ and $\bar{\Lambda}$ hyperons has
been measured in the STAR experiment, and the average angular
velocity or vorticity was estimated as $\omega\sim10^{22}s^{-1}$
\citep{STAR:2017ckg}. The data on the global polarization of $\Lambda$
and $\bar{\Lambda}$ hyperons can be well described by various phenomenological
models. 

The STAR Collaboration has also measured the local spin polarization
effect in Au+Au collisions at $200$ GeV \citep{Adam:2019srw}. The
data show a decreasing trend from in-plane to out-of-plane for the
global spin polarization. In addition, the STAR Collaboration measured the azimuthal angle
dependence of the spin polarization along the beam direction. These experimental results are still difficult to understand in terms of phenomenological
models.

This paper is structured as follows. In Sec. \ref{subsec:Several-recent-theoretical},
we review recent progress in theoretical studies of the CME and related
effects in heavy-ion collisions. In Sec. \ref{subsec:CMEexp}, we
present experimental results on the CME and CMW. In Sec. \ref{sec:Quantum-kinetic-theory},
we give a brief overview of recent developments in the quantum kinetic
theory for massive fermions. In Sec. \ref{subsec:Several-theoretical-updates},
we discuss the decomposition of the OAM and spin from the total angular
momentum as well as the spin hydrodynamics. In Sec. \ref{subsec:Recent-experimental-results},
we discuss experimental results on the global and local polarization effects
and spin alignments of vector mesons. We summarize the paper
in Sec. \ref{sec:Summary}. 

\section{Chiral magnetic and related effects in heavy-ion collisions \label{sec:Chiral-Magnetic-effect}}

\subsection{Theoretical progress\label{subsec:Several-recent-theoretical}}

To study the magnetic-field-related effects in heavy-ion collisions, we need to know the magnetic
field as a function of time. The electromagnetic
field can be estimated using the Lienard--Wiechert potential \citep{Bloczynski:2012en,Deng:2012pc,Roy:2015coa,Li:2016tel}.
Although the peak value of the magnetic field can be as large as a few times
$m_{\pi}^{2}$, where $m_{\pi}$ is the mass of a pion, it decays
very rapidly in vacuum \citep{Kharzeev:2007jp}. Such a magnetic
field in vacuum could not provide sufficient time to generate the CME
and other chiral transport effects; thus, intermediate effects must be present
\citep{McLerran:2013hla,Gursoy:2014aka,Li:2016tel,Gursoy:2018yai}. 

Magnetohydrodynamics (MHD) is widely used in astrophysics to investigate the coupling between
a charged medium and a magnetic field. Ordinary MHD consists of the hydrodynamic conservation equations
coupled with Maxwell's equations. Very recently, several MHD
studies \citep{Pu:2016ayh,Roy:2015kma,Pu:2016bxy,Pu:2016rdq,Roy:2017xtz}
showed that the magnetic fields decay as $1/\tau$ in the infinite
electrical conductivity limit, where $\tau$ is the proper time. The analytic
solution for the anomalous MHD with the CME and chiral
anomaly in a Bjorken flow has been derived \citep{Siddique:2019gqh,Wang:2020qpx}. The magnetic fields decay approximately
as $\sim1/\tau$ or $\sim e^{-\sigma\tau}/\tau$, where $\sigma$ is
the electrical conductivity. For numerical simulations of the ideal
MHD, see Ref. \citep{Inghirami:2016iru,Inghirami:2019mkc}. 
Although the electrical conductivity of the QGP is not infinite, it is important to study 
the coupling between the electromagnetic field and QGP. The ratio
of the magnetic density to the initial fluid density can reportedly be greater than $1$ in event-by-event simulations of
$\sqrt{s_{NN}}=200 \textrm{GeV}$ Au+Au collisions \cite{Roy:2015coa}; i.e., 
the magnetic field may also affect other physical quantities \cite{Pu:2016ayh,Pang:2016yuh, Roy:2017xtz}. 
Further study of the relativistic MHD with a temperature-dependent electrical conductivity is needed.


The ideal MHD
framework has been extended to the second order using the self-consistent
Grad moment expansion \citep{Denicol:2018rbw,Denicol:2019iyh}. In this framework, the electromagnetic fields
are coupled with normal dissipative terms, such as the shear viscous tensor
and bulk viscous pressure. One can also use a similar approach
based on the kinetic theory of massless fermions \citep{Shi:2020qrx}. 

By comparing phenomenological studies and experimental data, one can
constrain the lifetime of the magnetic field, $t_{B}\simeq0.5-1\textrm{fm/c}$ \citep{Muller:2018ibh}
or $t_{B}=A/\sqrt{s_{NN}}$, where $A=115\pm16\textrm{GeV}\cdot\textrm{fm/c}$
\citep{Guo:2019joy}. 

It has been proposed that the magnetic field can be produced by a charged
rotating fluid \citep{Guo:2019joy}. By using Maxwell's equations
in a charged fluid, $\partial_{\mu}F^{\mu\nu}=j^{\nu}$, and assuming
the particle number density $n$ is homogeneous or changes very slowly,
the relationship between the vorticity and the magnetic
field in the local rest frame, $\omega=(\nabla^{2}B)/en$, can be obtained. By introducing
the average vorticity $\bar{\boldsymbol{\omega}}$, the average magnetic
field per transverse area can be expressed as $\bar{\boldsymbol{B}}\simeq\frac{e^{2}}{4\pi}An\bar{\boldsymbol{\omega}}$,
where $A$ is the transverse area of the vortex. In heavy-ion collisions,
the evolution of both the vorticity and the charge density at freeze-out
can be extracted from AMPT simulations. (The AMPT model is a very good tool for simulating the CME
\cite{Huang:2017pzx, Deng:2018dut,Zhao:2019crj,Huang:2019vfy, Magdy:2020wiu}; see Sec. \ref{subsec:CMEexp}.) For a transverse area
of the vorticity of $16\pi\textrm{fm}^{2}$, and assuming a centrality of 20--50\% 
for Au+Au collisions at 10--200 GeV at the RHIC, the magnetic field and
its evolution can be estimated. Then, the splitting in the polarization
of $\Lambda$ and $\bar{\Lambda}$ hyperons can be estimated from the averaged
magnetic field, $P_{\Lambda}-P_{\bar{\Lambda}}\simeq2|\mu_{\Lambda}|\bar{B}/T_{\textrm{fo}}$,
where $|\mu_{\Lambda}|=0.613e/(2M_{N})$, $M_{N}=938$ MeV, and $T_{\textrm{fo}}=155$
MeV. The numerical results are close to the STAR data \citep{STAR:2017ckg}.

Another theoretical model is the anomalous-viscous fluid dynamics (AVFD)
\citep{Jiang:2016wve,Shi:2017cpu,Shi:2017ucn}, which is the hydrodynamical
realization of the CME in relativistic heavy-ion collisions. The latest
event-by-event version of the AVFD includes the local charge conservation (LCC)
effect \citep{Schenke:2019ruo} and the introduction of particlization,
which may be the best way to quantify the LCC \citep{Oliinychenko:2019zfk}.
It also gives predictions for isobaric collisions \citep{Shi:2019wzi}.
For the event plane,  AVFD predicts $\zeta_{isobar}^{EP}\equiv\left.\gamma_{Ru-Zr}^{OS-SS}\right|_{EP}/\left.\delta_{Ru-Zr}^{OS-SS}\right|_{EP}\simeq-(0.41\pm0.27)$,
which yields $\left\langle \cos(2\Psi_{B}-2\Psi_{EP})\right\rangle \simeq-0.46$;
for the reaction plane, $\zeta_{isobar}^{RP}\equiv\left.\gamma_{Ru-Zr}^{OS-SS}\right|_{RP}/\left.\delta_{Ru-Zr}^{OS-SS}\right|_{RP}\simeq-(0.90\pm0.45)$,
which yields $\left\langle \cos(2\Psi_{B}-2\Psi_{EP})\right\rangle \simeq-0.95$.
According to the calculations, these ratios are independent of the initial
axial charge. 


As a natural extension to the CME, it is interesting to consider how
large the mass correction is. This question can be addressed using a perturbation
method in quantum kinetic theory, as described in Sec. \ref{sec:Quantum-kinetic-theory}.
However, as noted in Ref. \citep{Fukushima:2010vw},
the mass correction to the CME is related to another well-known
phenomenon, Schwinger pair production. The operator equation for the
CME with finite mass corrections is the axial Ward identity, 
\begin{equation}
\partial_{\mu}j_{5}^{\mu}=2im\overline{\psi}\gamma^{5}\psi-\frac{e^{2}}{16\pi^{2}}\epsilon^{\mu\nu\alpha\beta}F_{\mu\nu}F_{\alpha\beta}.\label{eq:axial_ward}
\end{equation}
To simplify the discussion, one can assume that the electric and magnetic
fields $E$ and $B$, respectively, are constant and homogeneous in the $z$ direction.
In this case, the theory reduces to a (1+1)-dimensional problem; therefore,
it can be computed using the worldline formalism. However, the original
calculation by Schwinger \citep{Schwinger:1951nm} provides only the
value of $\left\langle \mathrm{out},0\left|\partial_{\mu}j_{5}^{\mu}\right|\mathrm{in},0\right\rangle =0$,
where $\left|\mathrm{in},0\right\rangle $ and $\left|\mathrm{out},0\right\rangle $
are the in and out states in vacuum, respectively. One needs to compute
the exception values of all the operators in (\ref{eq:axial_ward});
i.e., one needs to compute $\left\langle \mathrm{in},0\left|\partial_{\mu}j_{5}^{\mu}\right|\mathrm{in},0\right\rangle $.
The method of computing the Feynman propagator in the in-in state has
been used to obtain the following result by a long and technical calculation
\citep{Copinger:2018ftr}: 
\begin{equation}
\partial_{\mu}j_{5}^{\mu}=\frac{e^{2}EB}{2\pi^{2}}\exp\left(-\frac{\pi m^{2}}{eE}\right),
\end{equation}
which reduces to the standard chiral anomaly,
$\partial_{\mu}j_{5}^{\mu}=e^{2}EB/(2\pi^{2})$, in the massless limit. It is also consistent
with the physical scenario suggested in Ref. \citep{Fukushima:2010vw},
which implies that the chirality production rate should be proportional
to the Schwinger pair production rate. In addition, several other physical
quantities can also be obtained by the real-time worldline formalism.
The mass-corrected CME current is given by 
\begin{equation}
j^{z}=\frac{e^{2}EB}{2\pi^{2}}\coth\left(\frac{B}{E}\pi\right)\exp\left(-\frac{\pi m^{2}}{eE}\right),
\end{equation}
where the factor $\coth\left(B\pi/E\right)$ represents the sum over
the Landau levels.

There are also many studies of the anomaly and CME in the context of perturbation in quantum field theory
\cite{Feng:2017dom, Wu:2016dam, Lin:2018aon, Horvath:2019dvl, Feng:2018tpb}. The source of the CME, the chiral charge fluctuation \cite{Hou:2017szz, Lin:2018nxj}, is another research topic and is important for determining
the magnitude of the CME signal.

\subsection{Recent experimental results \label{subsec:CMEexp}}

The dipole charge separation due to the CME can be characterized by
the first sine term $a_{1}$ in the Fourier series of the charged-particle
azimuthal distribution:

\begin{equation}
\frac{dN}{d\phi}\propto1+2\sum_{n}\bigl\{ v_{n}\cos[n(\phi-\Psi_{\mathrm{RP}})]+a_{n}\sin[n(\phi-\Psi_{\mathrm{RP}})]\bigr\},\label{eq:azimuthal}
\end{equation}
where $\phi-\Psi_{\mathrm{RP}}$ represents the particle azimuthal
angle with respect to the reaction plane angle $\Psi_{\mathrm{RP}}$
in heavy-ion collisions (which is determined by the impact parameter and beam
axis), and $v_{n}$ and $a_{n}$ are the coefficients of the $P$-even
and $P$-odd Fourier terms, respectively. An azimuthal three-particle
correlator \citep{Voloshin:2004vk}, $\gamma_{112}$, was proposed to
explore the first coefficient, $a_{1}$, of the $P$-odd Fourier terms
characterizing the charge separation due to the CME; it is given by
\begin{equation}
\gamma_{112}\equiv\left\langle \cos(\phi_{\alpha}+\phi_{\beta}-2\Psi_{RP})\right\rangle \simeq\left\langle \cos(\phi_{\alpha}+\phi_{\beta}-2\Psi_{2})\right\rangle ,\label{eq:gamma-112}
\end{equation}
where $\alpha$ and $\beta$ denote particles with the same or opposite
electric charge, respectively; angle brackets denote the averages over particles
and events, and $\Psi_{RP}$ can be approximated by the
azimuthal angle of the second-order event plane, $\Psi_{2}$. Similarly, $\gamma_{123}$, 
a charge-dependent correlator with respect to the azimuthal
angle of the third-order event plane, $\Psi_{3}$, is defined
as 
\begin{equation}
\gamma_{123}\equiv\left\langle \cos(\phi_{\alpha}+2\phi_{\beta}-3\Psi_{3})\right\rangle \label{eq:gamma123}
\end{equation}
and can reflect the charge-dependent background effects unrelated
to the CME, because $\Psi_{3}$ is not correlated with 
the magnetic field direction. Other types of azimuthal correlators $\gamma_{ijk}$
for charged particles can also be defined similarly for different
purposes. Except in ambiguous cases, we use the shorthand notation $\gamma$
to represent $\gamma_{112}$ hereafter. 

\begin{figure}
\includegraphics[scale=0.45]{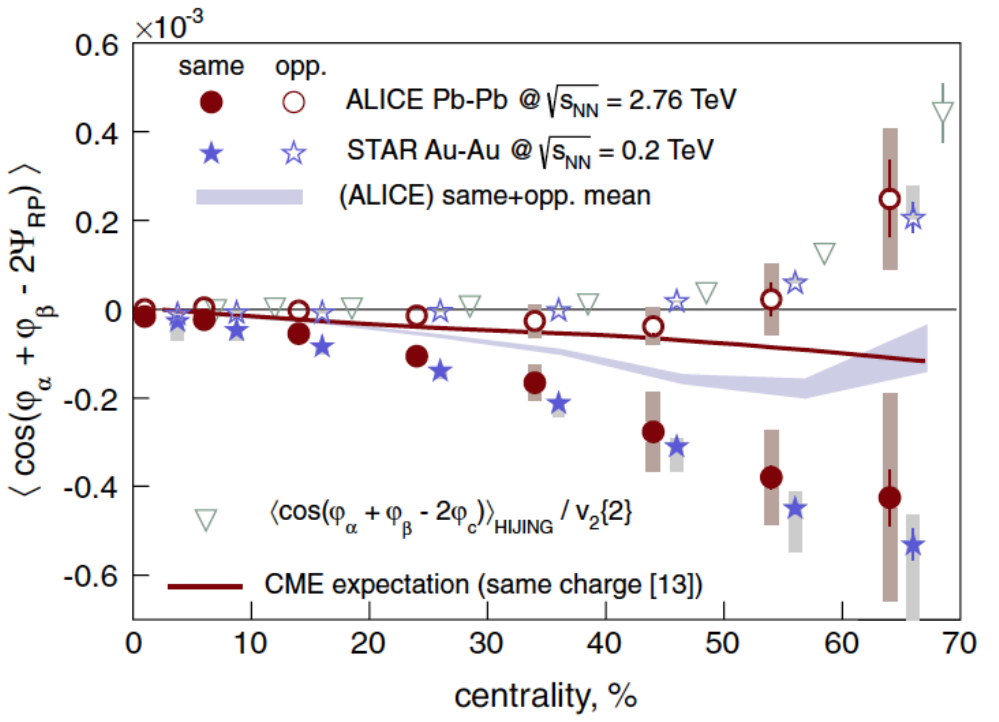}
\includegraphics[scale=0.57]{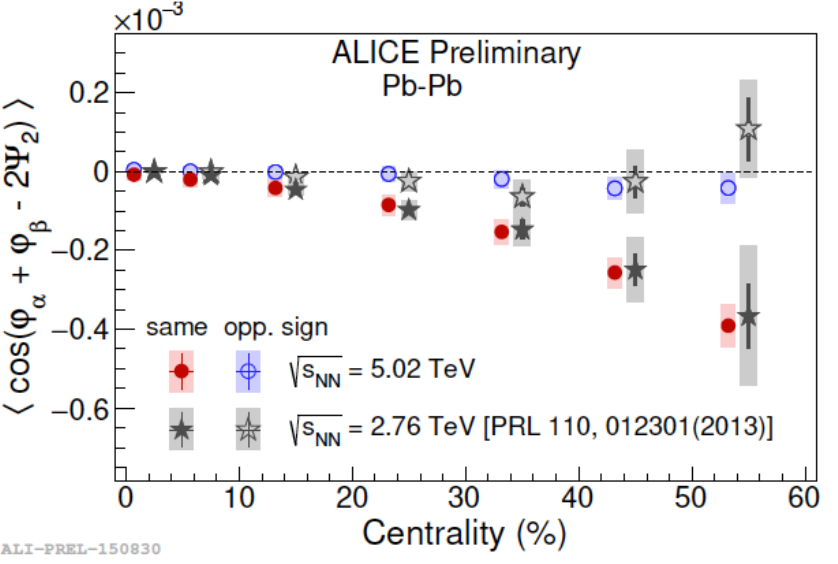} \caption{Measured centrality dependence of the $\gamma_{112}$ correlator
for Au+Au collisions at 200 GeV and Pb+Pb collisions at 2.76 TeV \citep{Abelev:2012pa}
(left panel) and for Pb+Pb collisions at 2.76 TeV and 5.02 TeV
(right panel) \citep{Aziz:2020qm19}. }
\label{FIGgammaAA}
\end{figure}

From the CME expectation, the difference between the opposite-sign
and same-sign correlation for charged particles $\Delta\gamma=\gamma_{{\rm opp}}-\gamma_{{\rm same}}$
is expected to be proportional to $B^{2}$ and ${\rm cos}\left[2(\Psi_{B}-\Psi_{2})\right]$
\citep{Bloczynski:2012en,Deng:2016knn,Zhao:2019crj}: 
\begin{equation}
\Delta\gamma\propto\left\langle B^{2}{\rm cos}\left[2(\Psi_{B}-\Psi_{2})\right]\right\rangle .\label{eq:delta}
\end{equation}
Because the magnitude of the magnetic field is proportional to the collision
energy \citep{Bzdak:2011yy,Deng:2012pc}, the CME should produce 
large differences between the $\gamma$ correlators at very different energies,
such as the RHIC and LHC energies. However, Fig. \ref{FIGgammaAA}
shows a very weak energy dependence of the $\gamma$ correlator in
a wide energy range from the RHIC energy to the LHC energy. 

\begin{figure}
\includegraphics[scale=0.6]{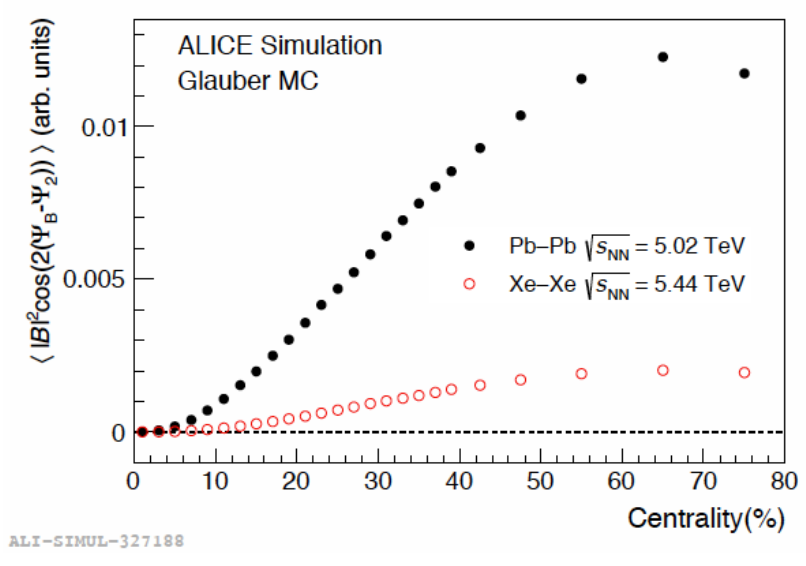} \includegraphics[scale=0.6]{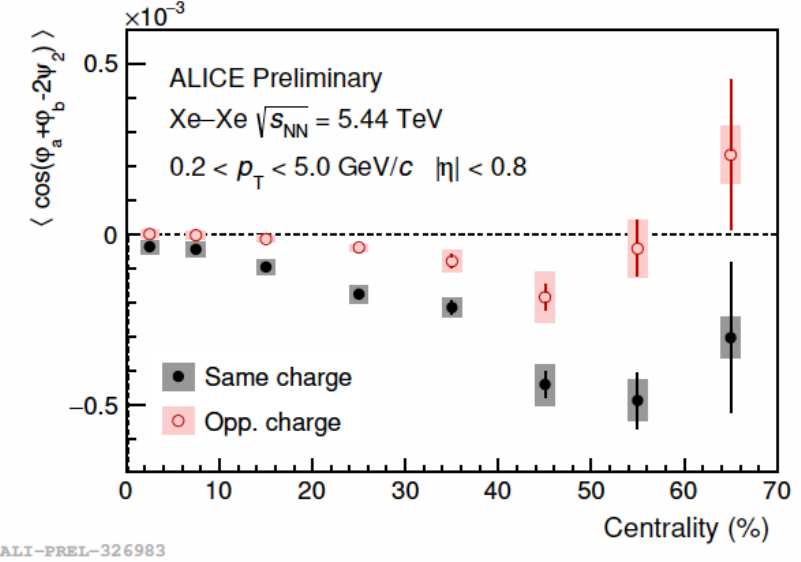}
\caption{Results of simulation of $\Delta\gamma$ as a function of centrality for
the CME in Pb+Pb collisions at 5.02 TeV and Xe+Xe collisions at 5.44
TeV (left panel); ALICE measurements of the $\gamma_{112}$ correlator
for Xe+Xe collisions at 5.44 TeV (right panel) \citep{Aziz:2020qm19}.}
\label{FIGgammaXeXe}
\end{figure}

As (\ref{eq:delta}) shows, the $\Delta\gamma$ resulting from the CME is proportional
to both the squared magnetic field and the correlation in the magnetic field direction with respect to the event plane. The left panel
of Fig. \ref{FIGgammaXeXe} shows the simulation results of $\Delta\gamma$
due to the CME as a function of the centrality for Pb+Pb collisions at 5.02
TeV and Xe+Xe collisions at 5.44 TeV. They are very different
except in the most central collisions. However, the right panel of Fig.
\ref{FIGgammaXeXe} shows the preliminary ALICE data for Xe+Xe collisions,
which show a centrality dependence that is very similar to that for Pb+Pb collisions
in Fig. \ref{FIGgammaAA}. The observed weak independence of  $\Delta\gamma$ from the collision energy indicates that the dominant component
of $\Delta\gamma$ is most likely due to the background.

\begin{figure}
\includegraphics[scale=0.5]{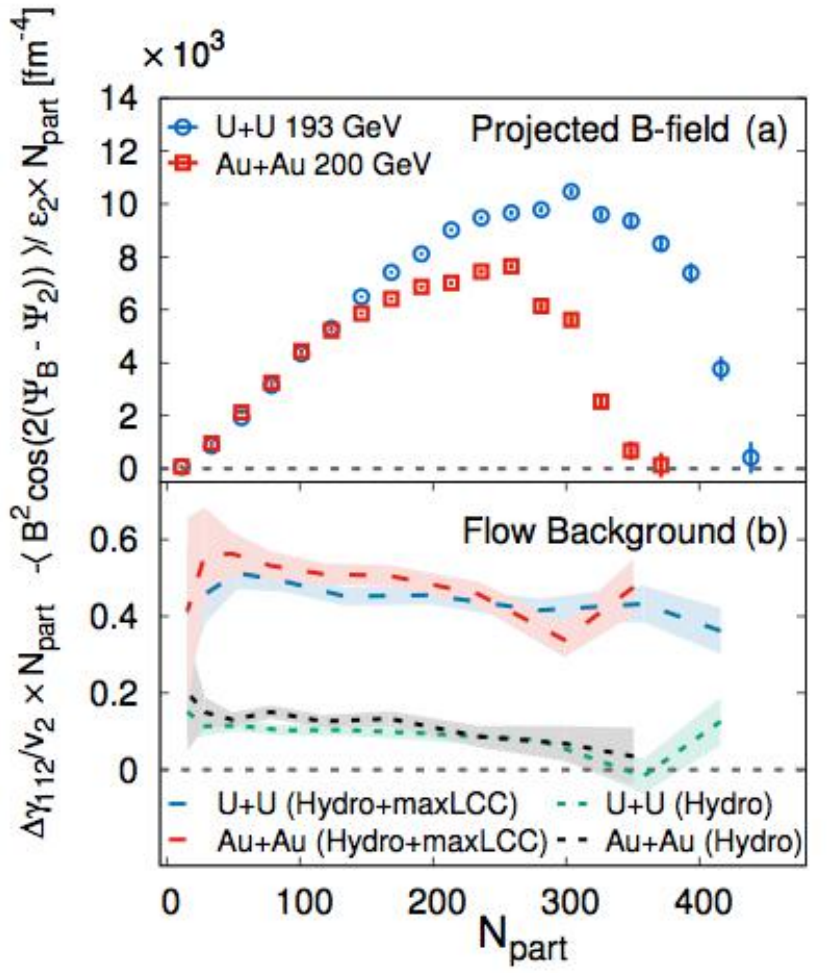} \includegraphics[scale=0.4]{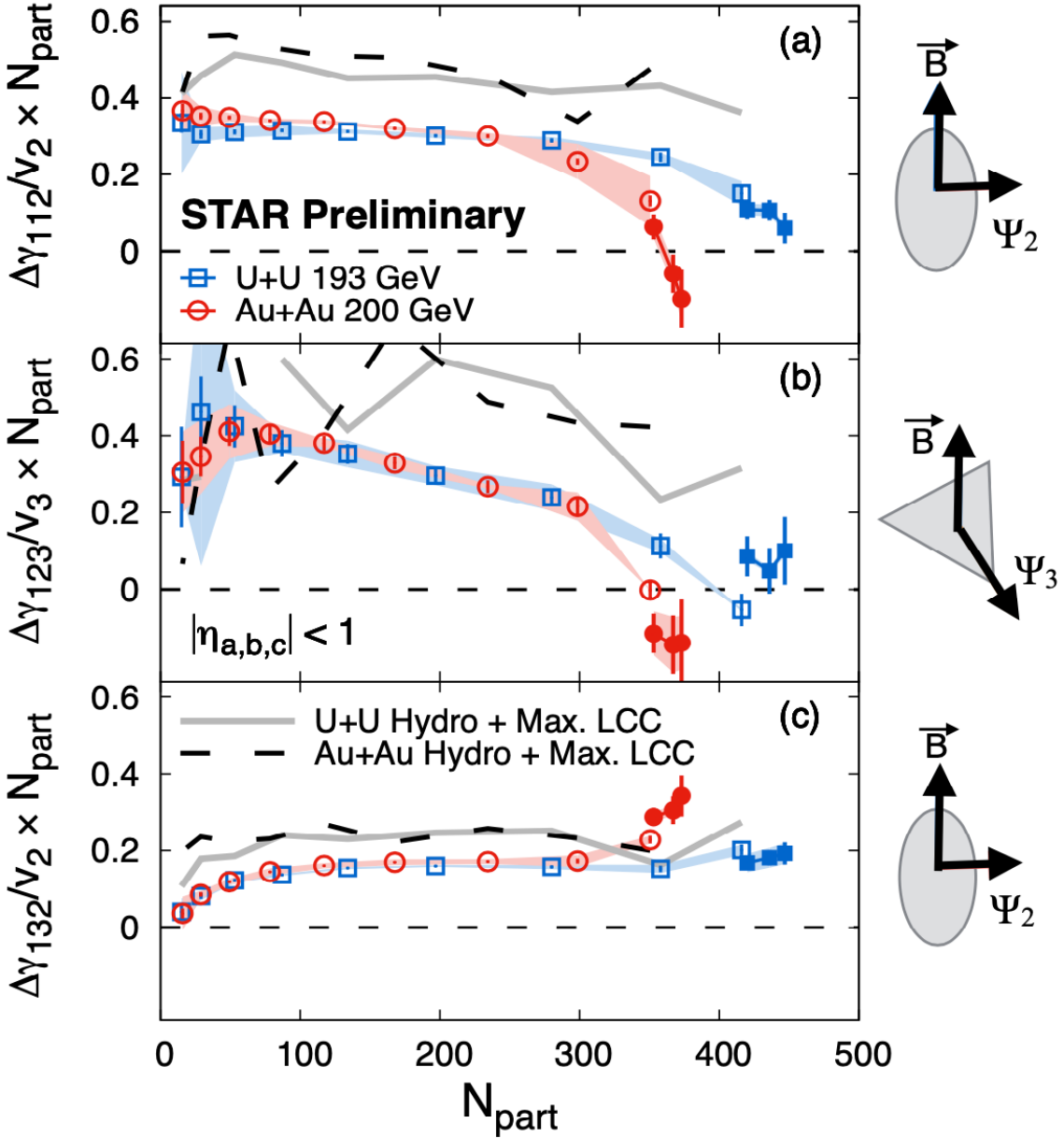}
\caption{ $N_{\mathrm{part}}$ dependence of the CME contribution (top left)
and flow background (bottom left) projected onto the correlator $\gamma_{112}$
for Au+Au collisions at 200 GeV and U+U collisions at 193 GeV. 
STAR measurements of $N_{\mathrm{part}}$ dependence of the correlators
($\Delta\gamma_{112}$, $\Delta\gamma_{123}$, and $\Delta\gamma_{132}$
multiplied by a factor of $N_{\mathrm{part}}/v_{n}$) for Au+Au collisions
at 200 GeV and U+U collisions at 193 GeV (right panel) \citep{Zhao:2020utk}. }
\label{FIGgammaUU}
\end{figure}

One may argue, however, that the CME signal also depends on the lifetime
of the magnetic field. Because of the shorter magnetic field lifetime
at higher collision energies, the signal is very weak at very high
energies. Fortunately, experimentalists are searching for a
possible signal by comparing Au+Au collisions and
U+U collisions at the RHIC. The top-left panel of Fig. \ref{FIGgammaUU}
shows the expected CME contribution to $\Delta\gamma$ as a function
of $N_{\mathrm{part}}$ for Au+Au collisions at 200 GeV and U+U collisions
at 193 GeV \citep{Zhao:2020utk}, which are multiplied by a factor
$N_{\mathrm{part}}/\epsilon_{n}$ to scale the effect from the elliptic
flow and transverse momentum conservation. The CME
contributions for Au+Au and U+U collisions are different except in
very peripheral collisions. The difference is sizable for $N_{\mathrm{part}}$
larger than 150. In addition, the background contribution to
$\Delta\gamma$ is simulated using a hydrodynamics model with and
without maximal LCC, as shown in the
bottom-left panel of Fig. \ref{FIGgammaUU}. For each
case, the expected contribution is almost the same for Au+Au collisions
at 200 GeV and U+U collisions at 193 GeV, although local charge
conservation can significantly increase the magnitude of $\Delta\gamma$.
To check these results, the STAR Collaboration has measured three
correlators, $\Delta\gamma_{112}$, $\Delta\gamma_{123}$, and $\Delta\gamma_{132}$,
in Au+Au collisions at 200 GeV and U+U collisions at 193 GeV \citep{Zhao:2020utk}.
Their $N_{\mathrm{part}}$ dependence is presented in the right
panel of Fig. \ref{FIGgammaUU}; they are normalized by a factor of
$N_{\mathrm{part}}/v_{n}$ for the above reason. It is observed
that the mixed harmonic correlations do not follow the background-only
expectations. Differences between these correlators in Au+Au collisions
and U+U collisions appear only for very central collisions with $N_{\mathrm{part}}$
larger than 300. The illustrations on the far right side of Fig.
\ref{FIGgammaUU} show that $\Delta\gamma_{123}$ is not
correlated with the magnetic field direction, in contrast to $\Delta\gamma_{112}$
and $\Delta\gamma_{132}$; it is hard to see whether this difference is due
to the signal or the background. A detailed study of these
correlators is needed to clarify their nature. 

\begin{figure}
\includegraphics[scale=0.4]{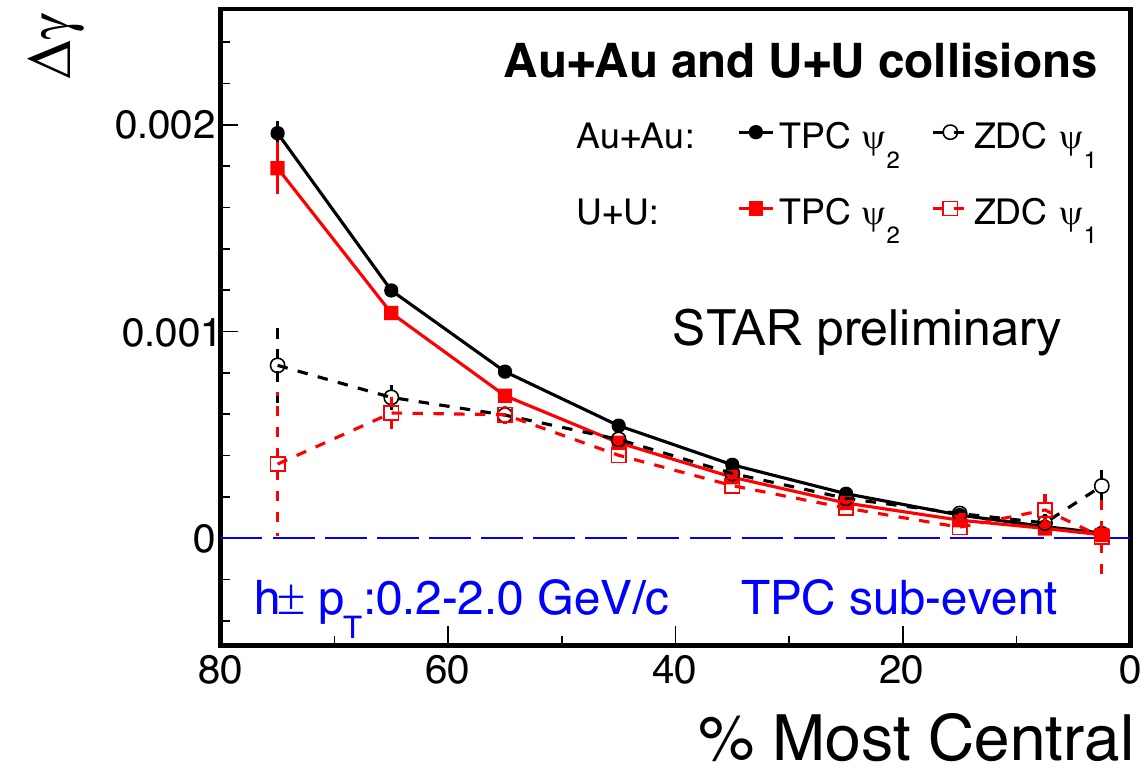} \includegraphics[scale=0.4]{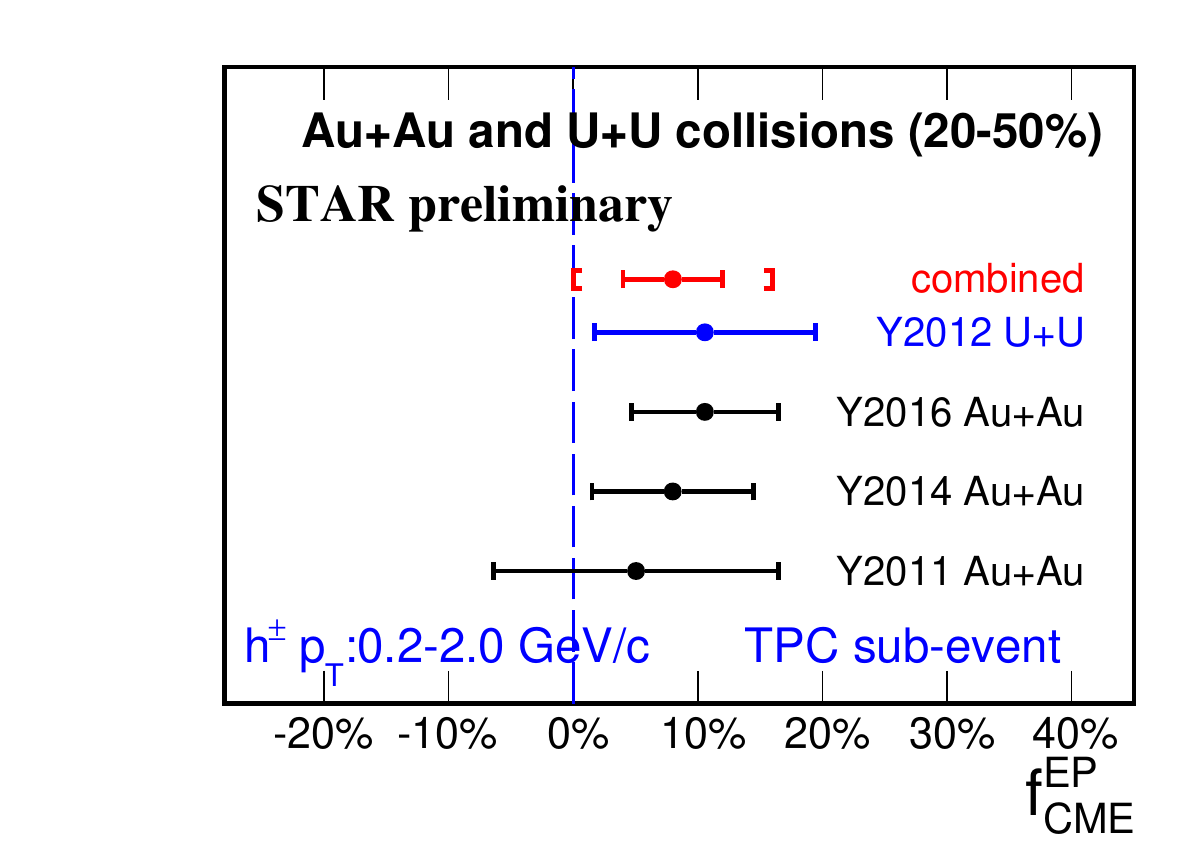}
\caption{ STAR measurements of the centrality dependence of $\Delta\gamma_{112}$
with respect to the ZDC and TPC event planes (left panel) and extracted
CME fractions $f_{\mathrm{CME}}$ (right panel) from Au+Au collisions
at 200 GeV and U+U collisions at 193 GeV \citep{Zhao:2020utk}. }
\label{FIGstarfCME}
\end{figure}

To extract the contribution of the CME to the observable
$\Delta\gamma$, the $\Delta\gamma$ measured in the STAR experiment was decomposed
into the $v_{2}$ background and the CME signal: 
\begin{eqnarray}
\Delta\gamma(\psi_{\mathrm{TPC}}) & = & \Delta\gamma_{\mathrm{CME}}(\psi_{\mathrm{TPC}})+\Delta\gamma_{\mathrm{Bkg}}(\psi_{\mathrm{TPC}}),\nonumber \\
\Delta\gamma(\psi_{\mathrm{ZDC}}) & = & \Delta\gamma_{\mathrm{CME}}(\psi_{\mathrm{ZDC}})+\Delta\gamma_{\mathrm{Bkg}}(\psi_{\mathrm{ZDC}}),\label{eq:tpc}
\end{eqnarray}
where $\psi_{\mathrm{TPC}}$ and $\psi_{\mathrm{ZDC}}$ are the event
planes measured for mid-rapidity particles in the time projection chamber (TPC)
and for spectator neutrons in the zero degree calorimeter (ZDC), respectively.
Assuming that the CME is proportional to the magnetic field squared and
the background is proportional to $v_{2}$ \citep{Xu:2017qfs}, one
obtains 
\begin{eqnarray}
\Delta\gamma_{\mathrm{CME}}(\psi_{\mathrm{TPC}}) & = & a\Delta\gamma_{\mathrm{CME}}(\psi_{\mathrm{ZDC}}),\nonumber \\
\Delta\gamma_{\mathrm{Bkg}}(\psi_{\mathrm{ZDC}}) & = & a\Delta\gamma_{\mathrm{Bkg}}(\psi_{\mathrm{TPC}}),
\end{eqnarray}
where $a=\left\langle \cos\left[2\left(\psi_{\mathrm{ZDC}}-\psi_{\mathrm{TPC}}\right)\right]\right\rangle $
and can be obtained from the $v_{2}$ measurement: 
\begin{equation}
a=\frac{v_{2}(\psi_{\mathrm{ZDC}})}{v_{2}(\psi_{\mathrm{TPC}})}.
\end{equation}
The CME signal relative to the inclusive $\Delta\gamma(\psi_{\mathrm{TPC}})$
can be determined as 
\begin{equation}
f_{\mathrm{CME}}=\frac{\Delta\gamma_{\mathrm{CME}}(\psi_{\mathrm{TPC}})}{\Delta\gamma(\psi_{\mathrm{TPC}})}=\frac{Aa-a^{2}}{1-a^{2}},\label{eqD}
\end{equation}
where $A$ is defined as 
\begin{eqnarray}
A & = & \frac{\Delta\gamma(\psi_{\mathrm{ZDC}})}{\Delta\gamma(\psi_{\mathrm{TPC}})}.
\end{eqnarray}
It can be rewritten as 
\begin{eqnarray}
A & = & f_{\mathrm{CME}}\frac{(1/a)\Delta\gamma_{\mathrm{CME}}(\psi_{\mathrm{TPC}})+a\Delta\gamma_{\mathrm{Bkg}}(\psi_{\mathrm{TPC}})}{\Delta\gamma_{\mathrm{CME}}(\psi_{\mathrm{TPC}})}\nonumber \\
 & = & a+\left(\frac{1}{a}-a\right)f_{\mathrm{CME}},
\end{eqnarray}
which one can solve for $f_{\mathrm{CME}}$ to obtain the last equality
of (\ref{eqD}). Note that $A$ can also be measured experimentally.
The left panel of Fig. \ref{FIGstarfCME} shows the STAR preliminary
data on the centrality dependence of $\Delta\gamma$ with respect to
the ZDC and TPC event planes in Au+Au collisions at 200 GeV and U+U
collisions at 193 GeV. We see that $\Delta\gamma(\psi_{\mathrm{ZDC}})$
is consistently lower than $\Delta\gamma(\psi_{\mathrm{TPC}})$, which
indicates that $A$ is less than unity. This result indicates that the flow background contributes less to $\Delta\gamma(\psi_{\mathrm{ZDC}})$
than to $\Delta\gamma(\psi_{\mathrm{TPC}})$.
Applying the above method, the STAR Collaboration extracted the CME
fraction $f_{\mathrm{CME}}$ from different datasets for Au+Au collisions
at 200 GeV and U+U collisions at 193 GeV; their results are summarized in
the right panel of Fig. \ref{FIGstarfCME}. The combined result for 
the CME fraction for Au+Au at 200 GeV and U+U at 193 GeV is $f_{{\rm CME}}=8\pm4\pm8\%$
\citep{Zhao:2020utk}. 
Note that the CMS Collaboration has systematically measured different types of correlators $\gamma_{ijk}$
in p+Pb collisions at 5.02 and 8.16 TeV and Pb+Pb collisions at 5.02 TeV, 
which provide constraints on the upper limit of the CME fraction, i.e., 13\% for p+Pb and 7\% for Pb+Pb collisions 
at the 95\% confidence level \citep{Khachatryan:2016got,Sirunyan:2017quh}.


\begin{figure}[!htb]
\includegraphics[scale=0.38]{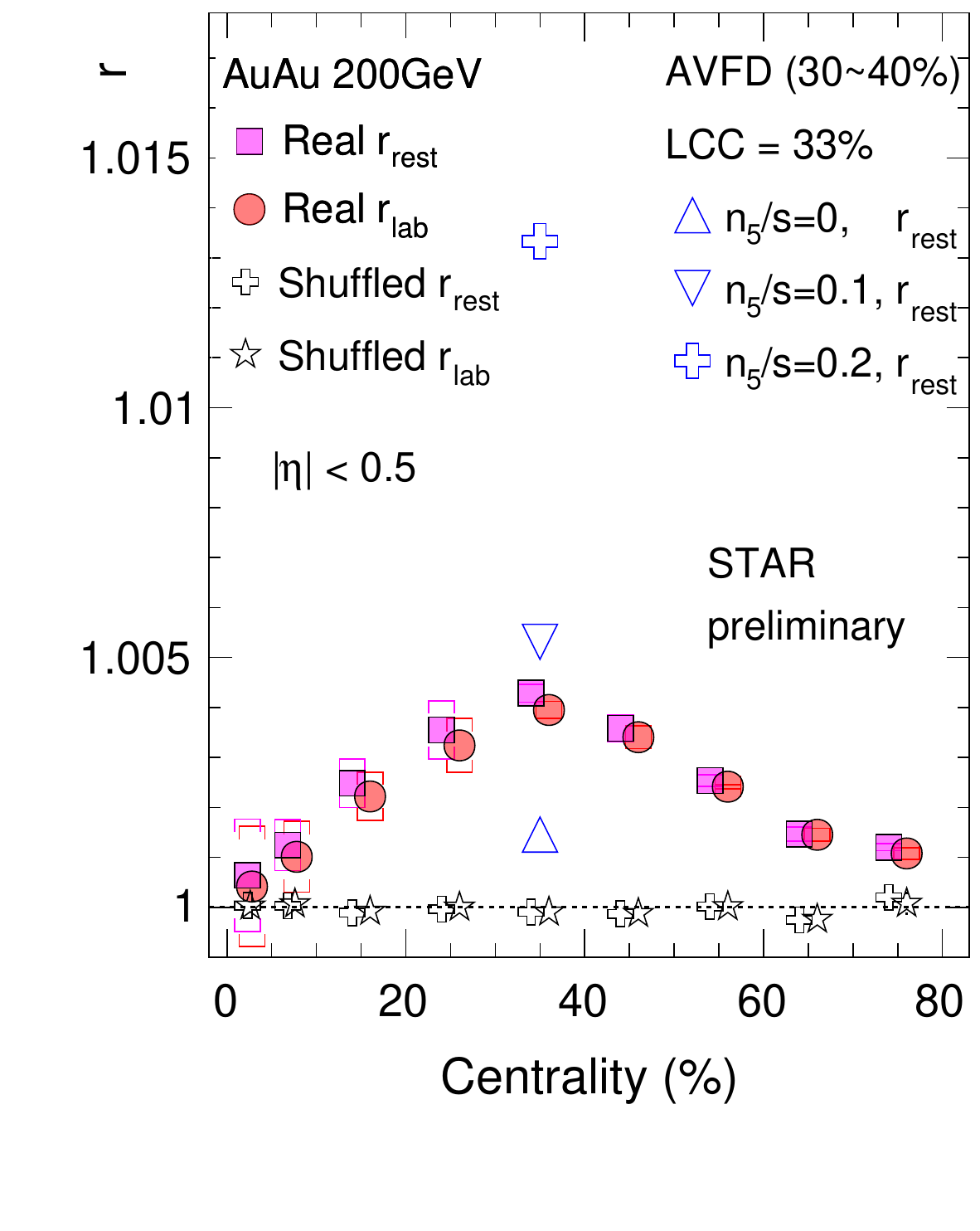} 
\includegraphics[scale=0.41]{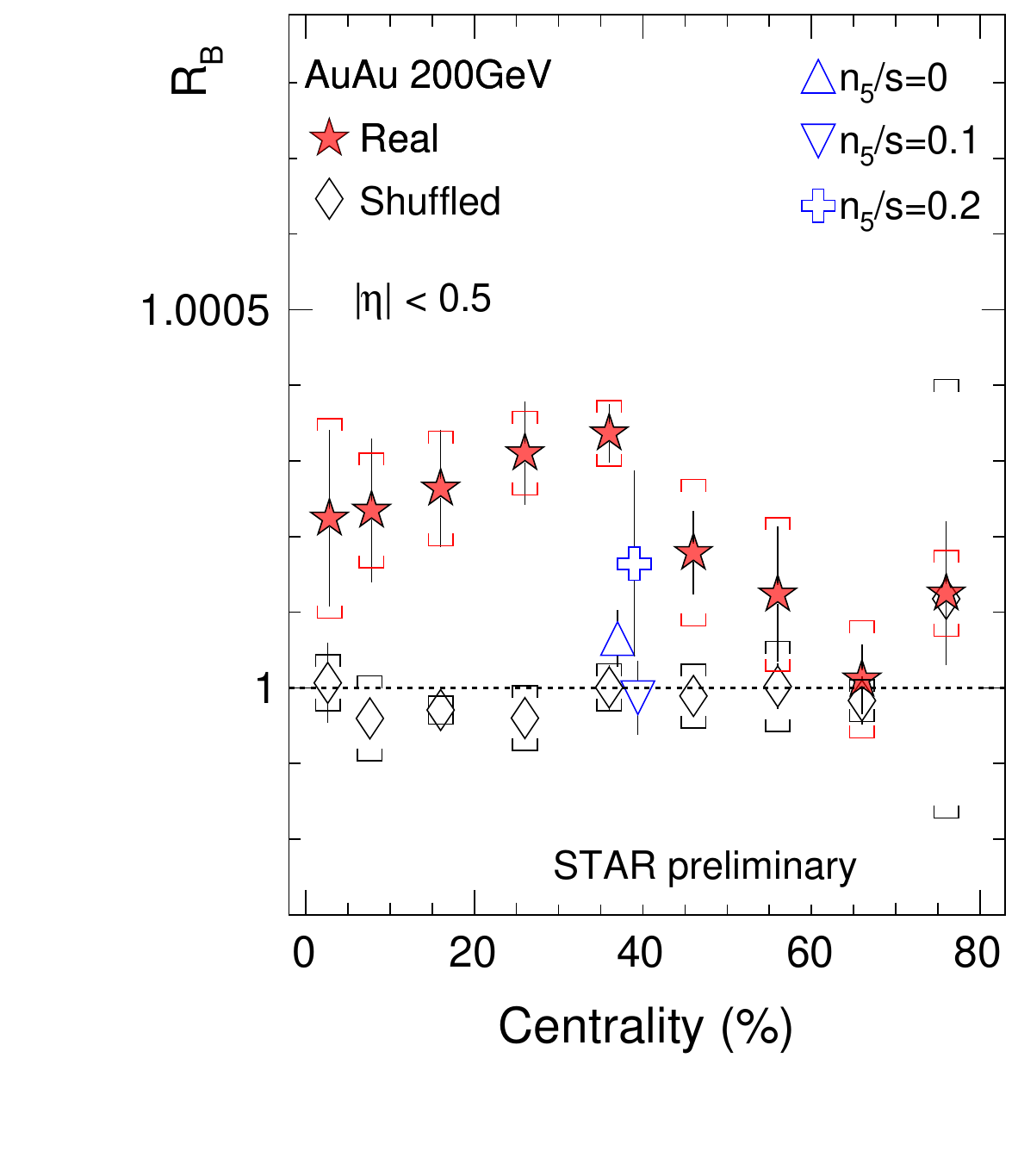}
\caption{ STAR measurements of centrality dependence of $r_{\mathrm{rest}}$,
$r_{\mathrm{lab}}$, and $R_{B}$ from signed balance functions in
Au+Au collisions at 200 GeV \citep{Lin:2020jcp}. }
\label{FIGsbfSTAR}
\end{figure}

Many other methods of searching for the CME have been proposed,
such as event shape engineering \citep{Sirunyan:2017quh}, the use of the H factor
\citep{Bzdak:2012ia}, and the invariant mass method \citep{Adam:2020zsu}. See
\citep{Kharzeev:2015znc,Li:2020dwr} for recent reviews. A new observable, the signed balance function, was proposed recently;
it is defined as \citep{Tang:2019pbl} 
\begin{eqnarray}
B_{P,y}(S_{y}) & = & \frac{N_{+-}(S_{y})-N_{++}(S_{y})}{N_{+}}\label{eq:Bp}
\end{eqnarray}
and 
\begin{eqnarray}
B_{N,y}(S_{y}) & = & \frac{N_{-+}(S_{y})-N_{--}(S_{y})}{N_{-}}.\label{eq:Bn}
\end{eqnarray}
Note that $S_{y}=+1$ if particle $\alpha$ is leading particle
$\beta$ (i.e., $p_{y}^{\alpha}>p_{y}^{\beta}$), and $S_{y}=-1$ otherwise. $N_{+-}(S_{y})$ denotes the number of positive--negative
charge pairs with sign $S_{y}$ in an event. $N_{++}(S_{y})$,
$N_{-+}(S_{y})$, and $N_{--}(S_{y})$ have similar definitions. $N_{+(-)}$
is the number of positive (negative) charge pairs in an event. Here
the $x$ axis is along the reaction plane, the $z$ axis
is along the beam direction, and the $y$ axis is perpendicular to
both the $x$ and $z$ axes. One can calculate an event using the
event difference between $B_{P}$ and $B_{N}$: 
\begin{eqnarray}
\delta B_{y}(\pm1) & = & B_{P,y}(\pm1)-B_{N,y}(\pm1)\label{eq:deltaB_pm}
\end{eqnarray}
and 
\begin{eqnarray}
\Delta B_{y} & = & \delta B_{y}(+1)-\delta B_{y}(-1).\label{eq:deltaB}
\end{eqnarray}
Note that $\Delta B_{x}$ can be defined similarly. When the
CME is absent, for a positive--negative charge pair, the probability
of the positive particle leading the negative one is equal to the probability
of the opposite case. Thus, $B_{P,y(x)}$ and $B_{N,y(x)}$
measure the same quantity, in principle, and the distribution of $\Delta B_{y(x)}$
is subject only to statistical fluctuation. When the CME is present,
the two probabilities become unbalanced within an event; thus, more pairs of particles of one charge type lead the other type.
Therefore, for each event, $B_{P,y}$ and $B_{N,y}$ tend to differ; consequently, $\Delta B_{y}$
has a wider distribution. By contrast, the distribution of $\Delta B_{x}$
is not broadened, as there is no charge separation in the $x$ direction.
To cancel out the statistical fluctuation, one can define the ratio
of the width of $\Delta B_{y}$ to that of $\Delta B_{x}$ as
\begin{equation}
r=\frac{\sigma_{\Delta B_{y}}}{\sigma_{\Delta B_{x}}}
\end{equation}
to characterize the magnitude of the CME, because $r$ will be greater
than unity or unity with or without the CME, respectively. That is, the strength of the CME will be positively correlated with
the deviation of $r$ from unity. The ratio $r$ can be calculated
either in the laboratory frame ($r_{\mathrm{lab}}$) or in the 
rest frame ($r_{\mathrm{rest}}$) of the pair. One can take the ratio of these two cases:
\begin{eqnarray}
R_{B} & \equiv & \frac{r_{\mathrm{rest}}}{r_{\mathrm{lab}}},\label{eq:R_B}
\end{eqnarray}
where the subscript $B$ indicates a balance function. It has been
found that $r_{\mathrm{lab}}$, $r_{\mathrm{rest}}$, and $R_{B}$
are sensitive not only to the strength of the CME, but also to the
elliptic flow of primordial pions and $\rho$ resonances, and even
to the global spin alignment of the resonances \citep{Tang:2019pbl}.
Figure \ref{FIGsbfSTAR} presents the recent STAR measurement of the centrality
dependence of $r_{\mathrm{lab}}$, $r_{\mathrm{rest}}$, and $R_{B}$
from the signed balance functions in Au+Au collisions at 200 GeV \citep{Lin:2020jcp}.
In the 30--40\% centrality bin, $r_{\mathrm{rest}}$,
$r_{\mathrm{lab}}$, and $R_{B}$ are all larger than the AVFD result
without the CME and larger than unity, which supports the presence of the
CME. However, none of the AVFD results for CMEs of different strengths can describe all three of these observables simultaneously. Therefore,
more experimental and theoretical studies are needed. 

\begin{figure}[!htb]
\includegraphics[scale=0.38]{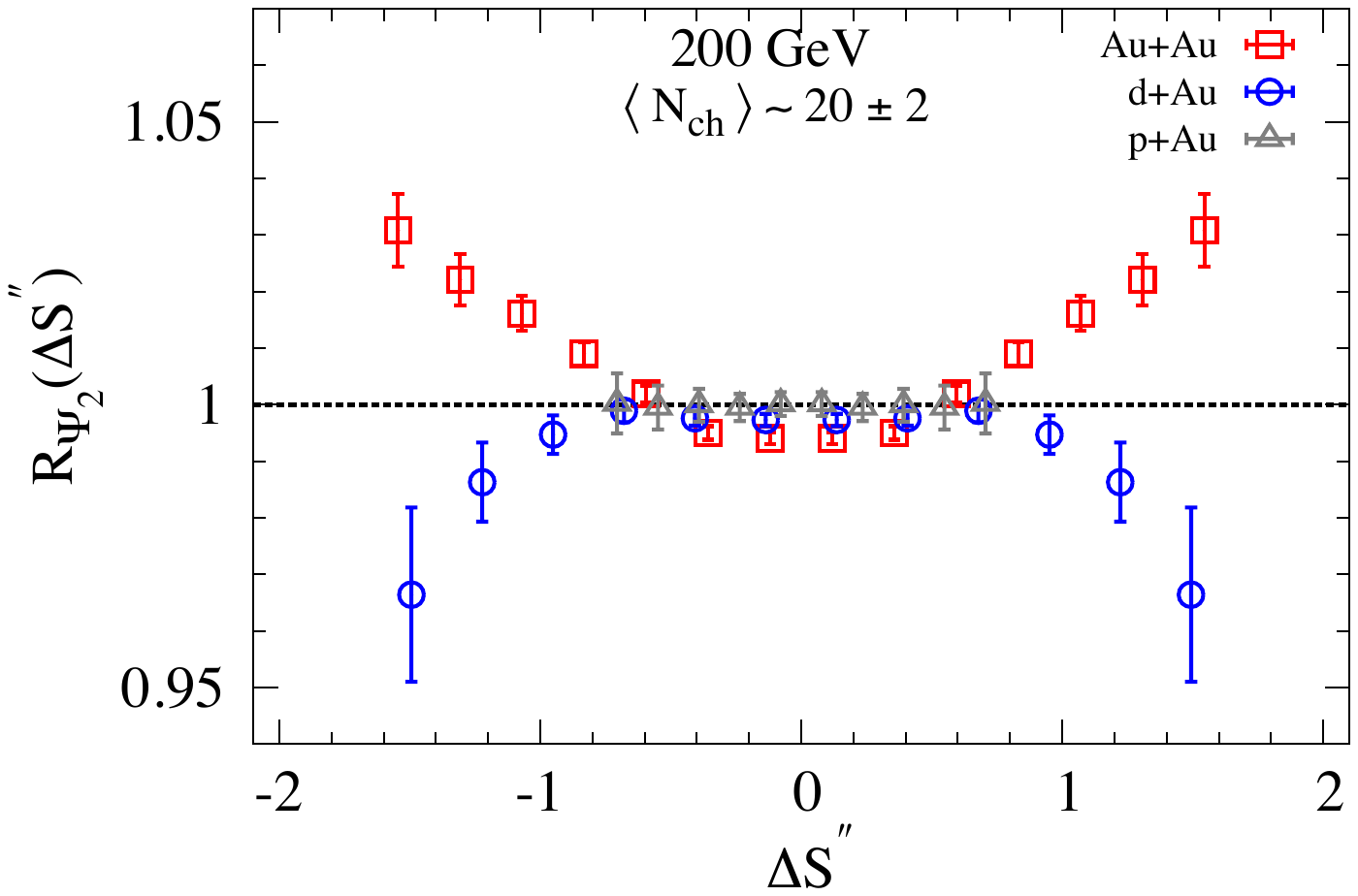} 
\caption{ STAR measurements of the $R_{\Psi_{2}}(\Delta S^{''})$ correlators obtained for charged particles in peripheral Au+Au and 
central $p$($d$)+Au collisions (${\left< N_{\rm ch} \right> \sim 20}$) at 200~GeV, where $\Delta S^{''}$ is the event-plane-resolution-corrected $\Delta S$ \citep{STAR:2020crk}. }
\label{FIGRcorrSTAR}
\end{figure}

Another new correlator, $R_{\Psi_2}(\Delta S) = C_{\Psi_2}(\Delta S)/C_{\Psi_2}^{\perp}(\Delta S)$, has been designed to measure the magnitude 
of the CME-induced charge separation parallel to the $\vec{B}$ field \citep{Magdy:2017yje,Magdy:2018lwk},
where $C_{\Psi_2}(\Delta S)$ and $C_{\Psi_2}^{\perp}(\Delta S)$ 
are used to quantify the charge separation $\Delta S$, i.e., the difference 
between the averaged $\left< \sin (\Delta {\varphi_{2}})\right>$ values of negatively and positively charged particles emitted 
about the estimated $\Psi_{2}$ plane, $\Delta {\varphi_{2}}= \phi - \Psi_{2}$.
Because the CME can result in charge separation along the $\vec{B}$ field, one expects a concave 
$R_{\Psi_2}(\Delta S)$ distribution having a width that reflects the magnitude $a_1$
of the CME-induced charge separation. Because the reaction plane angle $\Psi_2$ is very weakly correlated with the $\vec{B}$ field
 in small collision systems \citep{Zhao:2017rpf}, the corresponding measurement in small systems can provide a baseline 
 for the background contributions. The STAR Collaboration recently released the results for the $R$ correlator 
\citep{STAR:2020crk}. Figure~\ref{FIGRcorrSTAR} compares the $R_{\Psi_2}(\Delta S^{''})$ correlators 
at $\left<{N_{\rm ch}}\right> \sim 20$ for p+Au, d+Au, and Au+Au collisions at 200 GeV. The convex or flat distributions
shown for p(d)+Au collisions are consistent with the approximately 
random $\vec{B}$ field relative to $\Psi_2$ expected in small collisions.
By contrast, the Au+Au measurement relative to $\Psi_2$ shows a concave distribution, 
which is consistent with CME-driven charge separation according to the  
AVFD \citep{Magdy:2017yje} and AMPT models \citep{Huang:2019vfy,Magdy:2020wiu}.

\begin{figure}
\includegraphics[scale=0.8]{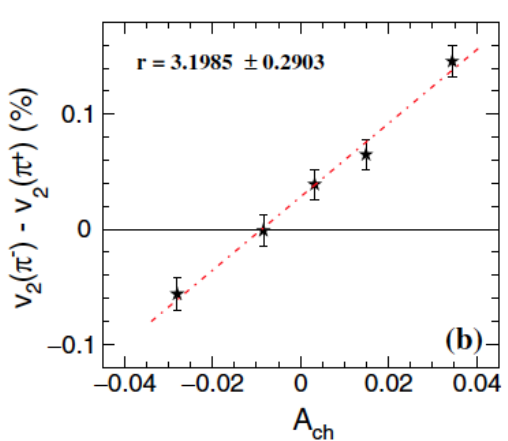} \includegraphics[scale=0.45]{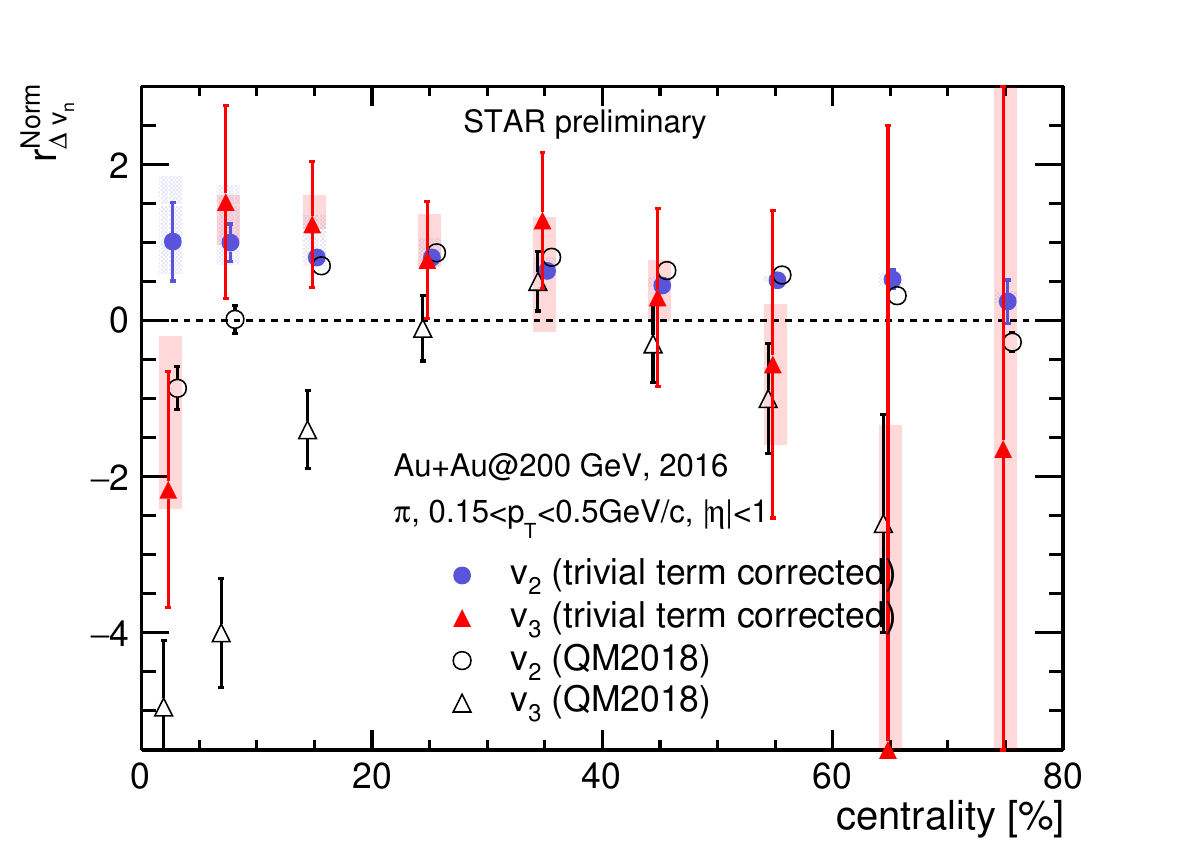}
\caption{STAR measurements of $v_{2}$ difference $\Delta v_{2}$ between
$\pi^{-}$ and $\pi^{+}$ as a function of charge asymmetry $A_{ch}$
for Au+Au collisions at 200 GeV (30--40\%) (left plot)~\citep{Adamczyk:2015eqo};
centrality dependence of the slope parameter $r$ of normalized
$\Delta v_{n}(A_{ch})$ with and without trivial term correction
in Au+Au collisions at 200 GeV~\citep{Xu:2017zcn, Xu:2020sln}.}
\label{FIGcmwSTAR}
\end{figure}

In addition to the CME, a gapless CMW could be formed
by the interplay between the CME and CSE. The CMW results in an electric
quadrupole moment in the initial coordinate space of the QGP
\citep{Kharzeev:2010gd,Burnier:2011bf}. It can ultimately be transformed 
into a charge-asymmetry-dependent elliptic flow of pions by collective
expansion \citep{Ma:2014iva}. Therefore, the elliptic flow of positively
and negatively charged pions is given by 
\begin{equation}
v_{2}(\pi^{\pm})=v_{2}^{{\rm base}}\mp\frac{r(\pi^{\pm})}{2}A_{\mathrm{ch}},
\end{equation}
where 
\begin{equation}
A_{\mathrm{ch}}=\frac{N^{+}-N^{-}}{N^{+}+N^{-}}
\end{equation}
is the charge asymmetry. The elliptic flow difference between
positive and negative pions, {[}$\Delta v_{2}=v_{2}(\pi^{-})-v_{2}(\pi^{+})${]},
can then be fitted using the relation $\Delta v_{2}=rA_{\mathrm{ch}}+\Delta v_{2}(0)$.
The left panel of Fig. \ref{FIGcmwSTAR} presents the results for Au+Au
collisions at 200 GeV (30--40\%) measured in the STAR experiment \citep{Adamczyk:2015eqo},
where the slope parameter $r$ is expected to reflect the strength
of the CMW. In addition, because the third order of the event plane
is not correlated with the magnetic field direction, measuring
the slope parameter $r$ from the triangular flow $v_{3}$ can provide
a reference from the background in comparison with that from elliptic
flow $v_{2}$. However, care should be taken, because flow measurements
by the Q-cumulant method using all the charged particles as a reference
can introduce a trivial linear term to $\Delta v_{n}(A_{{\rm ch}})$
owing to non-flow correlations. When all the charged hadrons are used as reference
particles, as is typically done in data analysis, the two-particle cumulant
can be rewritten as \citep{Xu:2019pgj} 
\begin{eqnarray}
d_{n}\{2;\pi^{\pm}h\} & = & \frac{d_{n}\{2;\pi^{\pm}h^{+}\}+d_{n}\{2;\pi^{\pm}h^{-}\}}{2}
 +\frac{d_{n}\{2;\pi^{\pm}h^{+}\}-d_{n}\{2;\pi^{\pm}h^{-}\}}{2}A_{\mathrm{ch}}.\label{eq:trivile}
\end{eqnarray}
The second term on the right-hand side of (\ref{eq:trivile})
is proportional to $A_{\mathrm{ch}}$ and is opposite in sign for $\pi^{+}$
and $\pi^{-}$. These characteristics result directly in a trivial contribution
to the CMW-sensitive slope parameter. Because non-flow correlations
are always present in experimental data, the flow coefficients
$d_{n}\{2;\mathrm{pairs}\}$ from like-sign pairs and opposite-sign
pairs differ. Therefore, the second term on the right-hand side of (\ref{eq:trivile})
is always finite as a trivial term, and it should be removed 
to detect the possible CMW signal. To eliminate the trivial
linear $A_{\mathrm{ch}}$ term in practice, one can use hadrons of
a single charge sign instead of all the charged hadrons as reference particles.
One may use positive and negative particles separately as reference particles
to obtain 
$v_{n}^{\pi}\{2;h^{+}\}$ and $v_{n}^{\pi}\{2;h^{-}\}$,
and then take an average: 
\begin{equation}
\bar{v}_{n}^{\pi}\equiv\frac{v_{n}^{\pi}\{2;h^{+}\}+v_{n}^{\pi}\{2;h^{-}\}}{2}.\label{eq:avgvn}
\end{equation}
Previous STAR results showed significant negative slopes for $v_{3}$
\citep{Shou:2018zvw}, which were thought to support the presence
of the CMW. By applying the new method of removing the trivial term,
the slope $r$ can be corrected to some extent. As shown in the right-hand
plot in Fig. \ref{FIGcmwSTAR}, the normalized $v_{3}$ slopes are
consistent with positive values (1.5$\sigma$ above zero for
20--60\% centrality), which are similar to the normalized $v_{2}$
slopes in terms of the relative magnitudes. In addition, it has been
found that the non-flow correlations give rise to additional background
in the slope of $\Delta v_{n}(A_{{\rm ch}})$ owing to competition
among different pion sources and the large multiplicity dilution
of $\pi^{+}$ ($\pi^{-}$) at positive (negative) $A_{{\rm ch}}$
\citep{Xu:2017zcn, Xu:2020sln}. Therefore, more detailed studies are needed
to search for the CMW in the future.

\section{Quantum kinetic theory \label{sec:Quantum-kinetic-theory}}

In recent years, chiral kinetic theory (CKT) has been developed significantly \citep{Stephanov:2012ki,Son:2012zy,Chen:2012ca,Manuel:2013zaa,Manuel:2014dza,Chen:2014cla,Chen:2015gta,Hidaka:2016yjf,Mueller:2017lzw,Hidaka:2017auj,Hidaka:2018mel,
Huang:2018wdl,Gao:2018wmr,Liu:2018xip,  Lin:2019ytz, Lin:2019fqo}
to describe various chiral effects in heavy-ion collisions.
Numerical simulations based on the CKT have been developed \citep{Sun:2016nig,Sun:2016mvh,Sun:2017xhx,Sun:2018idn,Sun:2018bjl,Zhou:2018rkh,Zhou:2019jag,Liu:2019krs}.
However, with the discovery of global polarization at relatively
low energies \citep{STAR:2017ckg,Adam:2018ivw}, it is necessary
to go beyond the chiral limit and develop a more general and practical
quantum kinetic theory to describe spin effects for massive fermions.
In this brief review, we consider only some of the works
that were reported at the 2019 Quark Matter conference. 

Most of these works are based on the Wigner functions but use slightly
different realizations. The methods used in Refs. \citep{Gao:2019znl,Weickgenannt:2019dks,Weickgenannt:2020sit}
are based on early works on the covariant Wigner functions \citep{Heinz:1983nx,Elze:1986qd,Vasak:1987um}.
The covariant Wigner function $W(x,p)$ for the Dirac fermion is defined
as a two-point function:
\begin{widetext}
\begin{equation}
W_{\alpha\beta}(x,p)=\int\frac{d^{4}y}{(2\pi)^{4}}e^{-ip\cdot y}\left\langle \bar{\psi}_{\beta}\left(x+\frac{y}{2}\right)U\left(x+\frac{y}{2},x-\frac{y}{2}\right)\psi_{\alpha}\left(x-\frac{y}{2}\right)\right\rangle ,\label{wigner}
\end{equation}
\end{widetext}
where $U$ denotes the gauge link along the straight line between
$x-y/2$ and $x+y/2$. The Wigner function can generally be expanded
as 
\begin{eqnarray}
W & = & \frac{1}{4}\left[\mathscr{F}+i\gamma^{5}\mathscr{P}+\gamma^{\mu}\mathscr{V}_{\mu}+\gamma^{5}\gamma^{\mu}\mathscr{A}_{\mu}+\frac{1}{2}\sigma^{\mu\nu}\mathscr{S}_{\mu\nu}\right]\;,\label{decomposition}
\end{eqnarray}
where $\sigma^{\mu\nu}=(i/2)[\gamma^{\mu},\gamma^{\nu}]$. One can
choose the scalar function $\mathcal{F}$ and axial vector function
$\mathcal{A}_{\mu}$ as independent variables \citep{Gao:2019znl};
they are related to $\mathscr{F}$ and $\mathscr{A}_{\mu}$ as 
\begin{eqnarray}
\mathscr{F} & = & \delta\left(p^{2}-m^{2}\right)\mathcal{F}+\frac{\hbar}{m}\tilde{F}_{\mu\nu}p^{\mu}\mathcal{A}^{\nu}\delta^{\prime}\left(p^{2}-m^{2}\right),\label{F-1-b}\\
\mathscr{A}_{\mu} & = & \delta\left(p^{2}-m^{2}\right)\mathcal{A}_{\mu}+\frac{\hbar}{m}p^{\nu}\tilde{F}_{\mu\nu}\mathcal{F}\delta^{\prime}\left(p^{2}-m^{2}\right).\label{A-1-b}
\end{eqnarray}
The kinetic equations for $\mathcal{F}$ and $\mathcal{A}_{\mu}$
are given by
\begin{eqnarray}
 &  & p\cdot\nabla\left[\mathcal{F}\delta\left(p^{2}-m^{2}\right)+\frac{\hbar}{m}\tilde{F}_{\mu\nu}p^{\mu}\mathcal{A}^{\nu}\delta^{\prime}\left(p^{2}-m^{2}\right)\right]\nonumber \\
 & = & \frac{\hbar}{2m}(\partial_{\lambda}^{x}\tilde{F}_{\mu\nu})\partial_{p}^{\lambda}\left[p^{\mu}\mathcal{A}^{\nu}\delta\left(p^{2}-m^{2}\right)\right]
\end{eqnarray}
and 
\begin{eqnarray}
 &  & p\cdot\nabla\left[\mathcal{A}_{\mu}\delta\left(p^{2}-m^{2}\right)+\frac{\hbar}{m}p^{\nu}\tilde{F}_{\mu\nu}\mathcal{F}\delta^{\prime}\left(p^{2}-m^{2}\right)\right]\nonumber \\
 & = & F_{\mu\nu}\left[\mathcal{A}^{\nu}\delta\left(p^{2}-m^{2}\right)+\frac{\hbar}{m}p_{\lambda}\tilde{F}^{\nu\lambda}\mathcal{F}\delta^{\prime}\left(p^{2}-m^{2}\right)\right]\nonumber \\
 &  & +\frac{\hbar}{2m}(\partial_{\lambda}^{x}\tilde{F}_{\mu\nu})\partial_{p}^{\lambda}\left[p^{\nu}\mathcal{F}\delta\left(p^{2}-m^{2}\right)\right],
\end{eqnarray}
together with the subsidiary condition $p\cdot\mathcal{A}\delta\left(p^{2}-m^{2}\right)=0$,
where $\nabla^{\mu}=\partial_{x}^{\mu}-F^{\mu\nu}\partial_{\nu}^{p}$,
and $\tilde{F}_{\mu\nu}=\epsilon_{\mu\nu\rho\sigma}F^{\rho\sigma}/2$.
The integrated kinetic equation in three-vector form can be obtained by
integrating the zero component of the momentum: 
\begin{widetext}
\begin{eqnarray}
\left(\nabla_{t}+\mathbf{v}\cdot\nabla\right)\mathcal{F} & = & -\frac{\hbar}{2mE_{p}}\left[({\bf B}+{\bf E}\times{\bf v})({\bf v}\cdot\nabla+E_{p}\overleftarrow{\nabla}_{x}\cdot\nabla_{p})\right.
 -\left.({\bf B}\cdot{\bf v})({\bf v}\cdot\nabla+E_{p}\overleftarrow{\nabla}_{x}\cdot\nabla_{p}{\bf v}\right]\cdot\overrightarrow{\mathcal{A}},\\
\left(\nabla_{t}+\mathbf{v}\cdot\nabla\right)\overrightarrow{\mathcal{A}} & = & {\bf B}\times\overrightarrow{\mathcal{A}}-{\bf E}({\bf v}\cdot\overrightarrow{\mathcal{A}})
-\frac{\hbar\,}{2mE_{p}}({\bf B}+{\bf E}\times{\bf v})({\bf v}\cdot\nabla+E_{p}\overleftarrow{\nabla}_{x}\cdot\nabla_{p})\mathcal{F},
\end{eqnarray}
\end{widetext}
where ${\bf v}={\bf p}/E_{p}$, $E_{p}=\sqrt{{\bf p}^{2}+m^{2}}$,
$\nabla_{t}=\partial_{t}+{\bf E}\cdot\nabla_{p}$, $\nabla=\nabla_{x}+{\bf B}\times\nabla_{p}$,
and the left arrow over $\nabla_{x}$ denotes that it acts only on
the electromagnetic fields on its left. When this method is used, various
spin effects such as the chiral anomaly, CSE, quantum magnetization
effect, and global polarization effect emerge naturally. It
can be shown that the Dirac sea or vacuum contribution originating
from the anti-commutation relations between antiparticle field operators
in the Wigner function without a normal-order operator plays a crucial role
in generating the chiral anomaly in quantum kinetic theory for both
massive and massless fermions \citep{Gao:2019zhk}. The coefficient
of the chiral anomaly derived in this way is universal and is independent
of the phase space distribution function at zero momentum.

Weickgenannt et al. \citep{Weickgenannt:2019dks,Weickgenannt:2020sit}
also derived the kinetic theory for massive spin-1/2 particles in the
covariant Wigner function formalism. They chose the scalar and tensor
components, $\mathscr{F}$ and $\mathscr{S}_{\mu\nu}$, as the basis from which the 
other components can be derived; these components are related to the distribution
function $V$ and tensor distribution or dipole moment $\bar{\Sigma}_{\mu\nu}$
as independent variables of $O(\hbar)$: 
\begin{eqnarray}
\mathscr{F} & = & \delta\left(p^{2}-m^{2}\right)mV-\frac{\hbar}{2}mF_{\mu\nu}\bar{\Sigma}^{\mu\nu}\delta^{\prime}\left(p^{2}-m^{2}\right), \nonumber \\
& & \label{F-1-c}\\
\mathscr{S}_{\mu\nu} & = & \delta\left(p^{2}-m^{2}\right)m\bar{\Sigma}_{\mu\nu}-\hbar F_{\mu\nu}V\delta^{\prime}\left(p^{2}-m^{2}\right), \nonumber \\
& &\label{S-1-c}
\end{eqnarray}
where $\delta^{\prime}(x)\equiv d\delta(x)/dx$, $V=V_{(0)}+\hbar V_{(1)}$,
and $\bar{\Sigma}^{\mu\nu}=\Sigma_{(0)}^{\mu\nu}A_{(0)}+\hbar\Sigma_{(1)}^{\mu\nu}$.
In addition, $V_{(0)}$ and $A_{(0)}$ are scalar functions of the phase space
and can be expressed as particle number distributions \citep{Weickgenannt:2019dks,Weickgenannt:2020sit}.
The kinetic equations for these functions are given by 
\begin{eqnarray}
0 & = & \delta(p^{2}-m^{2})\left[p\cdot\nabla V+\frac{\hbar}{4}(\partial_{x}^{\alpha}F^{\mu\nu})\partial_{\alpha}^{p}\bar{\Sigma}_{\mu\nu}\right]\nonumber \\
 &  & -\frac{\hbar}{2}\delta^{\prime}(p^{2}-m^{2})F^{\alpha\beta}\,p\cdot\nabla\bar{\Sigma}_{\alpha\beta}\nonumber \\
0 & = & \delta(p^{2}-m^{2})\bigg[p\cdot\nabla\bar{\Sigma}_{\mu\nu}-F_{\ [\mu}^{\alpha}\bar{\Sigma}_{\nu]\alpha}+\frac{\hbar}{2}(\partial_{x}^{\alpha}F_{\mu\nu})\partial_{p}^{\alpha}V\bigg]\nonumber \\
 &  & -\hbar\delta^{\prime}(p^{2}-m^{2})F_{\mu\nu}\,p\cdot\nabla V.\label{eq:kin-VandSigmaA}
\end{eqnarray}
together with one constraint equation, $p^{\nu}\bar{\Sigma}_{\mu\nu}\,\delta(p^{2}-m^{2})=\hbar\delta(p^{2}-m^{2})\nabla_{\mu}^V/2$.
In this method, the kinetic equations and the components of the Wigner
function are invariant under the transformation 
\begin{eqnarray}
\bar{\Sigma}_{\mu\nu} & \rightarrow & \widehat{\bar{\Sigma}}_{\mu\nu}=\bar{\Sigma}_{\mu\nu}+(p^{2}-m^{2})\delta\bar{\Sigma}_{\mu\nu}\,,\nonumber \\
V & \rightarrow & \widehat{V}=V-\frac{\hbar}{2}F^{\mu\nu}\delta\bar{\Sigma}_{\mu\nu}\,,
\end{eqnarray}
or the transformation 
\begin{eqnarray}
V & \rightarrow & \widehat{V}=V+(p^{2}-m^{2})\delta V\,,\nonumber \\
\bar{\Sigma}_{\mu\nu} & \rightarrow & \widehat{\bar{\Sigma}}_{\mu\nu}=\bar{\Sigma}_{\mu\nu}-\hbar F_{\mu\nu}\delta V\,,
\end{eqnarray}
where $\delta V$ and $\delta\bar{\Sigma}_{\mu\nu}$ are arbitrary
functions of $x$ and $p$ that are nonsingular on the mass shell.
When these transformations are used, the derivative
terms of the delta function can be omitted from the kinetic equations, greatly reducing the
results. Agreement with the CKT has
been obtained by replacing the dipole-moment tensor $\bar{\Sigma}_{\mu\nu}$
with its counterpart in the massless case. The kinetic equations yield a
special single-particle distribution in global equilibrium
with rigid rotation. When this distribution is used, analogs of various
spin effects appear, such as the CME and CSE, for massive
fermions. A smooth connection between the kinetic theory of 
massive spin-$1/2$ particles and the CKT was found recently \cite{Sheng:2020oqs}
by introducing a frame-dependent decomposition of the dipole-moment tensor
$\bar\Sigma_{\mu\nu}$.

All the components of the covariant Wigner function can be derived from
$\mathscr{V}^{\mu}$ and $\mathscr{A}^{\mu}$ for massive fermions
in the Schwinger--Keldysh formalism \citep{Hattori:2019ahi}. The CKT
for the massless case can be recovered from the formula for massive fermions.
The kinetic equation has been generalized to include collision terms
in the Schwinger--Keldysh formalism \citep{Yang:2020hri}. 

A covariant kinetic equation for massive fermions
in curved spacetime and an external electromagnetic field can also be derived. Liu et al. \citep{Liu:2020flb}
use a general covariant Wigner function formalism not
only under the $U(1)$ gauge and local Lorentz transformation but also
under diffeomorphism, which is compatible with general relativity.
The spin polarization in the presence of Riemann curvature and an electromagnetic
field in both local and global equilibrium has also been studied.

Integration of the covariant Wigner function over $p_{0}$
yields the equal-time Wigner function \citep{BialynickiBirula:1991tx,Zhuang:1998bqx,Gorbar:2017awz}.
Compared to the covariant form, the equal-time form loses obvious
Lorentz covariance, but it is more convenient for time evolution
problems such as pair production in a strong electric field \citep{Hebenstreit:2010vz,Sheng:2018jwf}.
The kinetic equation can also be derived from the equal-time Wigner functions.
The equal-time Wigner function $\mathcal{W}(x,{\bf p})$ can be obtained
from the covariant one $W(x,p)$: 
\begin{equation}
\mathcal{W}(x,{\bf p})=\int dp_{0}W(x,p)\gamma^{0}.
\end{equation}
It can be decomposed into the following components: 
\begin{eqnarray}
\mathcal{W} & = & \frac{1}{4}\left[f_{0}+\gamma_{5}f_{1}-i\gamma_{0}\gamma_{5}f_{2}+\gamma_{0}f_{3}+\gamma_{5}\gamma_{0}{\bf \gamma}\cdot{\bf g}_{0}\right.\nonumber \\
 &  & \left.+\gamma_{0}{\bf \gamma}\cdot{\bf g}_{1}-i{\bf \gamma}\cdot{\bf g}_{2}-\gamma_{5}{\bf \gamma}\cdot{\bf g}_{3}\right].
\end{eqnarray}
The fermion number density $f_{0}$ and spin current ${\bf g}_{0}$
are chosen to be independent components; their kinetic equations
are derived as \citep{Wang:2019moi}
\begin{widetext}
\begin{eqnarray}
\label{transport1-f}
\left(\nabla_t\pm\frac{{\bf p}}{E_p} \cdot {\pmb{ \nabla}}\right) f^{\pm}_0
&=& \frac{\hbar{\bf E}}{2E_p^2}\cdot{\pmb{ \nabla}}\times{\bf g}_0^{\pm}
\mp\frac{\hbar}{2E_p^3}{\bf B}\cdot({\bf p}\cdot{\pmb{ \nabla}}){\bf g}_0^{\pm}
+\frac{\hbar{\bf B}\times{\bf p}}{E_p^4}\cdot {\bf E}\times{\bf g}_0^{\pm},\\
\label{transport1-g0}
\left(\nabla_t\pm\frac{{\bf p}}{E_p}\cdot{\pmb{ \nabla}}\right){\bf g}^{\pm}_0
&=& {1 \over E_p^2}\left[{\bf p}\times \left({\bf E}\times {\bf g}^{\pm}_0\right)
\mp E_p{\bf B}\times{\bf g}^{\pm}_0\right]
\mp \hbar\left(\frac{ \bf B}{2E_p^3}\pm\frac{{\bf E}\times{\bf p}}{2E_p^4}\right){\bf p}\cdot{\pmb{ \nabla}} f^{\pm}_0\nonumber\\
&&\mp\hbar \left(\frac{({\bf p}\cdot{\bf E})({\bf E}\times{\bf p})}{E^5_p}\pm\frac{{\bf p}\times({\bf B}\times{\bf E})}{2E_p^4}\right)f^{\pm}_0.
\end{eqnarray}
\end{widetext}
where
the superscripts $+$ and
$-$ indicate the particle and antiparticle, respectively. The small
mass expansion can be performed. The mass correction
changes only the structure of the chiral kinetic equation in terms
of the effective collision terms, which is only a first-order quantum
correction to the CME.

All the works mentioned above are valid only up to the first order of $\hbar$ and
without collision terms for particle scattering. In addition to these works
based on Wigner functions, Li and Yee \citep{Li:2019qkf}
derived the kinetic equations for the spin-density matrix in the Schwinger--Keldysh
formalism. Specifically, they formulated collision terms for fermions
from their interactions with the QGP medium. However, they consider only the spatial homogeneity limit; thus, the collisions are local 
and do not involve the vorticity. The time evolution equations are
derived for the particle number distribution $f(\mathbf{p},t)$ and the
spin polarization density $\mathbf{S}(\mathbf{p},t)$ in the leading log order
of $g^{4}\log(1/g)$: 
\begin{eqnarray}
\frac{\partial f(\mathbf{p},t)}{\partial t} & = & C_{2}(F)\frac{m_{D}^{2}g^{2}\log(1/g)}{(4\pi)}\cdot\frac{1}{2pE_{p}}\Gamma_{f}\,,\nonumber \\
\frac{\partial\mathbf{S}(\mathbf{p},t)}{\partial t} & = & C_{2}(F)\frac{m_{D}^{2}g^{2}\log(1/g)}{(4\pi)}\cdot\frac{1}{2pE_{p}}\,\mathbf{\Gamma}_{S}\,,
\end{eqnarray}
where $C_{2}(F)=(N_{c}^{2}-1)/2N_{c}$, and $\Gamma_{f}$ and $\mathbf{\Gamma}_{S}$
are diffusion-like differential operators in momentum space that contain
up to the second-order derivatives of the momentum. The explicit forms of
$\Gamma_{f}$ and $\mathbf{\Gamma}_{S}$ are given in Eq. (4.70) of Ref.
\citep{Li:2019qkf}. 

Note that the quantum kinetic equation for massive fermions is very different from the chiral kinetic equation for massless fermions. The kinetic equation for massive fermions includes four coupled  independent functions; one of them gives the particle distribution function, and the other three give the spin polarization vector. By contrast, the kinetic equation for massless fermions contains only one distribution function. The reason is that the spin direction coincides with the momentum of the  fermion and is not a dynamical quantity at the chiral limit; however, for the massive case, the spin does not coincide with the momentum and becomes dynamical. It was thought that the chiral kinetic equation could be obtained from the quantum kinetic equation for massive fermions by naively taking the massless limit; however, this limit was found to be more important than it appears.  The most recent works on this problem can be found
in Refs. \cite{Sheng:2020oqs, Guo:2020zpa}.

\section{Spin polarization effects \label{sec:Polarization-effects}}

\subsection{Theoretical progress \label{subsec:Several-theoretical-updates}}

In early works \citep{ZTL_XNW_2005PRL,ZTL_XNW_2005PLB}, Liang
and Wang proposed that particles can be polarized as a result of
the global OAM in non-central heavy-ion collisions.
The formation of the vorticity in heavy-ion collisions was subsequently studied
\citep{Betz:2007kg}. Becattini and his
collaborators \citep{Becattini:2007nd,Becattini:2007sr,Becattini:2013fla} performed a systematic study of statistical models of
relativistic spinning particles. 
In addition, recent review articles are cited in Refs. \cite{Wang:2017jpl, Becattini:2020ngo, Becattini:2020sww}. 

The initial global OAM is estimated to be as large
as $J_{0}\sim10^{5}\hbar$ in 200 GeV Au+Au collisions with the impact
parameter $b=10\textrm{fm}$. Consequently, the QGP is found to
be the most vortical fluid ever observed in nature. The vorticity properties can be studied using the AMPT and HIJING \citep{Jiang:2016woz,Deng_2016PRC,LiHui_prc2017_lpwx},
UrQMD \citep{WeiDexian_prc2019_wdh}, and hydrodynamic models \citep{Csernai:2013bqa,Becattini:2013vja,Becattini:2015ska,Pang:2016igs,Wu:2019eyi, Wu:2020yiz}.
The vorticity is found to be more strongly suppressed at higher collision
energies. One can find further discussion in Refs. \citep{Ivanov_prc2017_is,Ivanov_prc2018_is,Ivanov_prc2019_its}
and reference therein. 

The spin polarization per particle for spin-1/2 fermions with 
momentum $p$ at freeze-out can be derived using the statistical model
for relativistic spinning particles \citep{Becattini:2013fla} and
the Wigner functions \citep{Fang:2016uds}: 
\begin{widetext}
\begin{equation}
S^{\mu}(p)=-\frac{1}{8m}\frac{\epsilon^{\mu\nu\rho\sigma}p_{\nu}\int d\Sigma_{\lambda}p^{\lambda}(u\cdot p)^{-1}f_{\mathrm{FD}}(1-f_{\mathrm{FD}})\omega_{\rho\sigma}^{\mathrm{th}}(x)}{\int d\Sigma_{\lambda}p^{\lambda}f_{\mathrm{FD}}},\label{eq:PL_vector}
\end{equation}
\end{widetext}
where $\Sigma_{\lambda}$ denotes the freeze-out hypersurface, $S^{\mu}$
is the Pauli--Lubanski pseudo-vector, $m$ is the fermion mass, $f_{\mathrm{FD}}\equiv f_{\mathrm{FD}}(x,\mathbf{p})$
is the Fermi--Dirac distribution. In addition, 
\begin{equation}
\omega_{\rho\sigma}^{\mathrm{th}}=\frac{1}{2}\left(\partial_{\sigma}\beta_{\rho}-\partial_{\rho}\beta_{\sigma}\right)\label{eq:thermal-vort}
\end{equation}
is the thermal vorticity, where $u^{\mu}$ is the fluid velocity,
and $\beta_{\rho}=u_{\rho}/T$ is the temperature four-vector.
Equation (\ref{eq:PL_vector}) has been widely used in hydrodynamics
models and transport models to calculate the spin polarization of hyperons.

In quantum field theory, the decomposition
of the total angular momentum into the orbital and spin components for massive particles with spin
is not unique. 
Different decompositions are related by a pseudo-gauge
transformation. 
Although the energy-momentum tensor $\hat{T}^{\mu\nu}$
and spin tensor $\hat{S}^{\lambda;\mu\nu}$ depend on the 
decomposition, the total energy-momentum and total angular momentum
are invariant under the pseudo-gauge transformation. 
A special pseudo-gauge
transformation is the transformation of the canonical definitions of $\hat{T}_{C}^{\mu\nu}$
and $\hat{S}_{C}^{\lambda;\mu\nu}$ into the Belinfante $\hat{T}_{B}^{\mu\nu}$,
which absorbs $\hat{S}_{C}^{\lambda;\mu\nu}$; i.e., $\hat{S}_{B}^{\lambda;\mu\nu}=0$.
Thus, the total angular momentum consists only of the orbital component. The Belinfante
energy-momentum tensor is symmetric with respect to its two indices:
$\hat{T}_{B}^{\mu\nu}=\hat{T}_{B}^{\nu\mu}$. 

The thermal properties of a quantum field system can be described
by the density operator $\hat{\rho}$. The density operator in local
equilibrium can be determined from the maximal entropy principle \citep{Zubarev_tmp1979_zps,Weert_ap1982,Becattini:2014yxa}
under given densities of conserved currents on the space-like hypersurface
$\Sigma$. This is done by maximizing $S=-\mathrm{Tr}(\hat{\rho}\log\hat{\rho})$
under the conditions 
\begin{eqnarray}
n_{\mu}\mathrm{Tr}\left(\hat{\rho}\hat{T}^{\mu\nu}\right) & = & n_{\mu}T^{\mu\nu},\nonumber \\
n_{\mu}\mathrm{Tr}\left(\hat{\rho}\hat{j}^{\mu}\right) & = & n_{\mu}j^{\mu},\label{eq:condition-rho}
\end{eqnarray}
where $n_{\mu}$ is the normal vector of $\Sigma$. The constraint on the total angular momentum density can generally be included as well:
\begin{eqnarray}
n_{\lambda}\mathrm{Tr}\left(\hat{\rho}\hat{\mathcal{J}}^{\lambda;\mu\nu}\right) & = & n_{\lambda}\mathrm{Tr}\left[\hat{\rho}\left(x^{\mu}\hat{T}^{\lambda\nu}-x^{\nu}\hat{T}^{\lambda\mu}+\hat{S}^{\lambda;\mu\nu}\right)\right]\nonumber \\
 & = & n_{\lambda}\mathcal{J}^{\lambda;\mu\nu}.\label{eq:oam-constraint}
\end{eqnarray}
For the Belinfante energy-momentum tensor, however, the above condition is automatically
satisfied by using (\ref{eq:condition-rho}). In this case, the
density operator at local equilibrium has the form 
\begin{equation}
\hat{\rho}_{\mathrm{LE}}=\frac{1}{Z}\exp\left[-\int_{\Sigma}d\Sigma_{\mu}\left(\hat{T}_{B}^{\mu\nu}\beta_{\nu}-\zeta\hat{j}^{\mu}\right)\right],\label{eq:loca;-eq}
\end{equation}
where $\beta^{\nu}$ and $\zeta$ are the Lagrange multiplier
functions of space-time; the former is the temperature four-vector,
and the latter is the ratio of the chemical potential to the temperature
\citep{Becattini:2014yxa}. 

In general, $\hat{\rho}_{\mathrm{LE}}$ in (\ref{eq:loca;-eq}) depends
on the time $\tau$ as $\Sigma_{\mu}(\tau)$. One can rewrite
the exponent of $\hat{\rho}_{\mathrm{LE}}$ as 
\begin{eqnarray}
& &\int_{\Sigma(\tau)}d\Sigma_{\mu}\left(\hat{T}_{B}^{\mu\nu}\beta_{\nu}-\zeta\hat{j}^{\mu}\right)  \nonumber\\
& = & \int_{\Sigma(\tau_{0})}d\Sigma_{\mu}\left(\hat{T}_{B}^{\mu\nu}\beta_{\nu}-\zeta\hat{j}^{\mu}\right)\nonumber \\
 &   - &\int_{\Theta}d\Theta\left(\hat{T}_{B}^{\mu\nu}\partial_{\mu}\beta_{\nu}-\hat{j}^{\mu}\partial_{\mu}\zeta\right),
\end{eqnarray}
where $\Theta$ is the space-time volume bounded by $\Sigma(\tau)$,
$\Sigma(\tau_{0})$, and the time-like boundary connecting $\Sigma(\tau)$
and $\Sigma(\tau_{0})$. Here we used 
\begin{equation}
\partial_{\mu}\left(\hat{T}_{B}^{\mu\nu}\beta_{\nu}-\zeta\hat{j}^{\mu}\right)
 =  \hat{T}_{B}^{\mu\nu}\partial_{\mu}\beta_{\nu}-\hat{j}^{\mu}\partial_{\mu}\zeta,
\end{equation}
where we used the conservation of energy-momentum and 
current. To make $\hat{\rho}_{\mathrm{LE}}$ independent of the
choice of $\Sigma_{\mu}(\tau)$, i.e., to obtain the global equilibrium density
matrix $\hat{\rho}_{\mathrm{GE}}$, the following conditions must
be met: 
\begin{eqnarray}
\partial_{\mu}\beta_{\nu}+\partial_{\nu}\beta_{\mu} & = & 0,\nonumber \\
\partial_{\mu}\zeta & = & 0,\label{eq:killing-cond}
\end{eqnarray}
according to which $\zeta$ is constant and $\beta^{\mu}$ is
a Killing vector: 
\begin{eqnarray}
\beta^{\mu} & = & \omega_{\mathrm{th}}^{\mu\nu}x_{\nu}+b^{\mu},\label{eq:killing-sol}
\end{eqnarray}
where $\omega_{\mathrm{th}}^{\mu\nu}$ is a constant antisymmetric
tensor given by (\ref{eq:thermal-vort}), and $b^{\mu}$ is a constant
vector. The local equilibrium density operator in (\ref{eq:loca;-eq})
can be expressed in terms of the canonical energy-momentum and spin tensors:
\begin{eqnarray}
\hat{\rho}_{\mathrm{LE}} & = & \frac{1}{Z}\exp\left\{-\int_{\Sigma}d\Sigma_{\mu}\left[\hat{T}_{C}^{\mu\nu}\beta_{\nu}-\zeta\hat{j}^{\mu}\right.\right.-\frac{1}{2}\omega_{\lambda\nu}^{\mathrm{th}}\hat{S}_{C}^{\mu;\lambda\nu} \nonumber \\
 &  & \left.\left.-\frac{1}{2}\left(\partial_{\lambda}\beta_{\nu}+\partial_{\nu}\beta_{\lambda}\right)\left(\hat{S}_{C}^{\lambda;\mu\nu}+\hat{S}_{C}^{\nu;\mu\lambda}\right)\right]\right\}.\label{eq:le-can-1}
\end{eqnarray}
If we substitute (\ref{eq:killing-cond}) and (\ref{eq:killing-sol}) in
the above expression for $\hat{\rho}_{\mathrm{LE}}$, we obtain $\hat{\rho}_{\mathrm{GE}}$:
\begin{equation}
\hat{\rho}_{\mathrm{GE}}=\frac{1}{Z}\exp\left[-b_{\mu}\hat{P}^{\mu}+\frac{1}{2}\omega_{\lambda\nu}^{\mathrm{th}}\hat{J}^{\lambda\nu}+\zeta\hat{Q}\right],\label{eq:rho-global-e}
\end{equation}
where $\hat{P}^{\mu}$, $\hat{J}^{\lambda\nu}$, and $\hat{Q}$ are
given by 
\begin{eqnarray}
\hat{P}^{\nu} & = & \int_{\Sigma}d\Sigma_{\mu}\hat{T}^{\mu\nu},\nonumber \\
\hat{J}^{\lambda\nu} & = & \int_{\Sigma}d\Sigma_{\mu}\mathcal{J}^{\mu;\lambda\nu},\nonumber \\
\hat{Q} & = & \int_{\Sigma}d\Sigma_{\mu}\hat{j}^{\mu}.
\end{eqnarray}

If we work with the canonical energy-momentum and spin tensors, we have
to impose the constraint in (\ref{eq:oam-constraint}), which also constrains the spin tensor: 
\begin{equation}
n_{\lambda}\mathrm{Tr}\left(\hat{\rho}\hat{S}_{C}^{\lambda;\mu\nu}\right)=n_{\lambda}S_{C}^{\lambda;\mu\nu}.
\end{equation}
This introduces a spin chemical potential, the antisymmetric tensor $\Omega_{\lambda\nu}$,
as a Lagrange multiplier into the equation for $\hat{\rho}_{\mathrm{LE}}$:
\begin{equation}
\hat{\rho}_{\mathrm{LE}}=\frac{1}{Z}\exp\left[-\int_{\Sigma}d\Sigma_{\mu}\left(\hat{T}_{C}^{\mu\nu}\beta_{\nu}-\frac{1}{2}\Omega_{\lambda\nu}\hat{S}_{C}^{\mu;\lambda\nu}-\zeta\hat{j}^{\mu}\right)\right].\label{eq:le-canonical}
\end{equation}
Comparing (\ref{eq:le-canonical}) with (\ref{eq:le-can-1}), we see that
$\hat{\rho}_{\mathrm{LE}}$ is the same for the Belinfante and canonical
tensors under the following conditions: (a) the field $\beta^{\mu}$
is the same in both cases; (b) $\Omega_{\lambda\nu}=\omega_{\lambda\nu}^{\mathrm{th}}$; and
(c) $\left(\partial_{\lambda}\beta_{\nu}+\partial_{\nu}\beta_{\lambda}\right)\left(\hat{S}_{C}^{\lambda;\mu\nu}+\hat{S}_{C}^{\nu;\mu\lambda}\right)=0$.
We can see that if $\Omega_{\lambda\nu}=\omega_{\lambda\nu}^{\mathrm{th}}$
and $\partial_{\lambda}\beta_{\nu}+\partial_{\nu}\beta_{\lambda}=0$,
then $\hat{\rho}_{\mathrm{LE}}$ for the canonical tensors becomes the
same as $\hat{\rho}_{\mathrm{GE}}$ for the Belinfante tensor in (\ref{eq:rho-global-e}).
Therefore, one can conclude that the density operator in global equilibrium
is pseudo-gauge-invariant, but in local equilibrium, it depends on
the pseudo-gauge \citep{Becattini_plb2019_bfs}. 

If we choose the Belinfante tensors or $\hat{\rho}_{\mathrm{LE}}$
in (\ref{eq:loca;-eq}) or (\ref{eq:le-can-1}), the
spin relaxation time is microscopically small, and the value of the
spin potential agrees with the thermal vorticity almost immediately. If
we choose the canonical tensors or $\hat{\rho}_{\mathrm{LE}}$ in
(\ref{eq:le-canonical}) with the spin chemical potential, the spin
density slowly reaches global equilibrium, just as a conserved
charge density or energy density does, and finally the spin chemical
potential should converge to the thermal vorticity \citep{Becattini_plb2019_bfs}.

Similar problems involving the decomposition of the total angular momentum of quarks
and gluons in the proton spin (see Ref. \citep{Leader:2013jra} for
a review) have been widely debated in the QCD community for years.
Two well-known decomposition methods are that of Jaffe and Manohar and that of Ji, which are close to the canonical and Belinfante tensor forms,
respectively. A review article \citep{Fukushima:2020qta}
discusses the difference between the Belinfante and canonical
forms. For a chiral medium, the decomposition can be found in Ref. \citep{Fukushima:2018osn}.

If we choose a specific pseudo-gauge, for example, the canonical form that
has the spin tensor, the spin chemical potential
$\Omega_{\lambda\nu}$ must be introduced into $\hat{\rho}_{\mathrm{LE}}$, as shown in
 (\ref{eq:le-canonical}). Therefore, we have another conservation
equation in addition to the conservation of energy-momentum and charge: 
\begin{equation}
\partial_{\mu}j^{\mu}=0,\ \ \partial_{\mu}T^{\mu\nu}=0,\ \ \partial_{\lambda}S^{\lambda,\mu\nu}=T^{\nu\mu}-T^{\mu\nu},
\end{equation}
where the energy-momentum tensor, charge current, and spin tensor can
be obtained through $\hat{\rho}_{\mathrm{LE}}$: 
\begin{eqnarray}
T^{\mu\nu}&=&\text{tr}\left(\hat{\rho}_{\mathrm{LE}}\hat{T}^{\mu\nu}\right), \nonumber\\
j^{\mu}&=&\text{tr}\left(\hat{\rho}_{\mathrm{LE}}\hat{j}^{\mu}\right), \nonumber \\
S^{\mu,\nu\lambda}&=&\text{tr}\left(\hat{\rho}_{\mathrm{LE}}\hat{S}^{\mu,\nu\lambda}\right).
\end{eqnarray}
All the above quantities are functions of $\beta^{\mu}$, $\zeta$, and
$\Omega_{\lambda\nu}$. There are a total of 11 independent equations
and variables. These equations make up a full set of hydrodynamic equations
including the spin degrees of freedom. There have been a few attempts
in spin hydrodynamics \citep{Florkowski_prc2018_ffjs,Florkowski:2018myy,Becattini_plb2019_bfs,Florkowski:2018fap,Hattori:2019lfp,Bhadury:2020puc,Shi:2020qrx}. 
Another approach using Lagrangian techniques can also provide some information about the local equilibrium 
\cite{Montenegro:2017rbu, Montenegro:2017lvf,Montenegro:2018bcf, Montenegro:2020paq}.

Spin degrees of freedom can be introduced into the phase space distribution
for spin-1/2 fermions by generalizing the scalar function to a $2\times2$
Hermitian matrix \citep{Becattini:2013fla}: 
\begin{eqnarray}
f_{rs}^{+}(x,p) & = & \frac{1}{2m}\bar{u}_{r}(p)X^{+}u_{s}(p),\nonumber \\
f_{rs}^{-}(x,p) & = & -\frac{1}{2m}\bar{v}_{r}(p)X^{-}v_{s}(p),
\end{eqnarray}
where $r,s=\pm1$; $u_{r}$ and $v_{r}$ are Dirac spinors normalized
by $\bar{u}_{r}(p)u_{s}(p)=2m\delta_{rs}$ and $\bar{v}_{r}(p)v_{s}(p)=-2m\delta_{rs}$, respectively; and
$X^{\pm}$ are $4\times4$ matrices defined as 
\begin{equation}
X^{\pm}=\exp\left(-\beta_{\mu}p^{\mu}\pm\zeta\pm\frac{1}{4}\Omega_{\mu\nu}\sigma^{\mu\nu}\right).
\end{equation}
The Wigner functions in equilibrium for fermions and antifermions
are given by 
\begin{eqnarray}
W_{\text{eq}}^{+}(x,k) & = & \sum_{r,s}\int d^{4}p\theta(p_{0})\delta(p^{2}-m^{2})\delta^{(4)}(k-p)
 u_{r}(p)\bar{u}_{s}(p)f_{rs}^{+}(x,p)\nonumber \\
 & = & \frac{1}{2m}\int d^{4}p\theta(p_{0})\delta(p^{2}-m^{2})\delta^{(4)}(k-p) (p_{\mu}\gamma^{\mu}+m)X^{+}(p_{\mu}\gamma^{\mu}+m),\nonumber \\
W_{\text{eq}}^{-}(x,k) & = & -\sum_{r,s}\int d^{4}p\theta(-p_{0})\delta(p^{2}-m^{2})\delta^{(4)}(k+p)
 v_{r}(p)\bar{v}_{s}(p)f_{rs}^{-}(x,p)\nonumber \\
 & = & \frac{1}{2m}\int d^{4}p\theta(-p_{0})\delta(p^{2}-m^{2})\delta^{(4)}(k-p)
(p_{\mu}\gamma^{\mu}-m)X^{-}(p_{\mu}\gamma^{\mu}-m).\label{eq:wigner-f-eq-f-af}
\end{eqnarray}
The total Wigner function is the sum of them: 
\begin{equation}
W_{\text{eq}}(x,k)=W_{\text{eq}}^{+}(x,k)+W_{\text{eq}}^{-}(x,k).\label{eq:wigner-function-eq-x}
\end{equation}
The components of $W_{\text{eq}}(x,k)$ are given in (\ref{decomposition})
and can be extracted by taking the traces of $\Gamma_{i}W_{\text{eq}}(x,k)$
with $\Gamma_{i}=1,\gamma^{5},\gamma^{\mu},\gamma^{5}\gamma^{\mu},\sigma^{\mu\nu}$.

The energy-momentum and spin tensors proposed by de Groot, van Leeuwen,
and van Weert (GLW) \citep{DeGroot:1980dk} can be extracted from
the Wigner function \citep{Florkowski:2018fap}:
\begin{eqnarray}
T_{\mathrm{GLW}}^{\mu\nu} & = & \frac{1}{m}\int d^{4}kk^{\mu}k^{\nu}\mathscr{F},\nonumber \\
S_{\mathrm{GLW}}^{\lambda;\mu\nu} & = & \frac{1}{4}\int d^{4}k\mathrm{Tr}\left[\left\{ \sigma^{\mu\nu},\gamma^{\lambda}\right\} W+\frac{2i}{m}\left(\gamma^{[\mu}k^{\nu]}\gamma^{\lambda}-\gamma^{\lambda}\gamma^{[\mu}k^{\nu]}\right)W\right].
\end{eqnarray}
In equilibrium, we use $W=W_{\mathrm{eq}}$ in (\ref{eq:wigner-function-eq-x})
and obtain $T_{\mathrm{GLW}}^{\mu\nu}=T_{\mathrm{eq}}^{\mu\nu}$ and
$S_{\mathrm{GLW}}^{\lambda;\mu\nu}=S_{\mathrm{eq}}^{\lambda;\mu\nu}$.
Because the GLW energy-momentum tensor is symmetric, the GLW spin tensor
should be conserved separately. Thus, the conservation laws read 
\begin{equation}
\partial_{\mu}j^{\mu}=0,\ \ \partial_{\mu}T_{\mathrm{GLW}}^{\mu\nu}=0,\ \ \partial_{\lambda}S_{\mathrm{GLW}}^{\lambda;\mu\nu}=0.\label{eq:cons-glw}
\end{equation}

The canonical forms of the energy-momentum and spin tensors can be
obtained from the Dirac Lagrangian \citep{Vasak:1987um,Weickgenannt:2019dks}:
\begin{eqnarray}
T_{C}^{\mu\nu} & = & \int d^{4}kk^{\nu}\mathscr{V}^{\mu},\nonumber \\
S_{C}^{\lambda;\mu\nu} & = & \frac{1}{4}\int d^{4}k\mathrm{Tr}\left[\left\{ \sigma^{\mu\nu},\gamma^{\lambda}\right\} W\right]
 = -\frac{1}{2}\epsilon^{\lambda\mu\nu\rho}\int d^{4}k\mathscr{A}_{\rho}.
\end{eqnarray}
One can verify that the canonical tensor forms are related to the GLW forms
by the pseudo-gauge transformation \citep{Florkowski:2018fap}. The explicit relation between the two tensor forms can be written as
\begin{eqnarray}
T_{C}^{\mu\nu} & = & T_{\mathrm{GLW}}^{\mu\nu}-\frac{\hbar}{2m}\int d^{4}kk^{\nu}\partial_{\lambda}S_{\mathrm{eq}}^{\lambda\mu}= T_{\mathrm{GLW}}^{\mu\nu}-\partial_{\lambda}S_{\mathrm{GLW}}^{\nu;\lambda\mu},\nonumber \\
S_{C}^{\lambda;\mu\nu} & = & S_{\mathrm{GLW}}^{\lambda;\mu\nu}+S_{\mathrm{GLW}}^{\mu;\nu\lambda}+S_{\mathrm{GLW}}^{\nu;\lambda\mu}.
\end{eqnarray}
One can verify that the conservation laws for $T_{C}^{\mu\nu}$ and
$S_{C}^{\lambda;\mu\nu}$ follow those for $T_{\mathrm{GLW}}^{\mu\nu}$
and $S_{\mathrm{GLW}}^{\lambda;\mu\nu}$ in (\ref{eq:cons-glw}).

\begin{figure}[tb]
\begin{centering}
\includegraphics[scale=0.5]{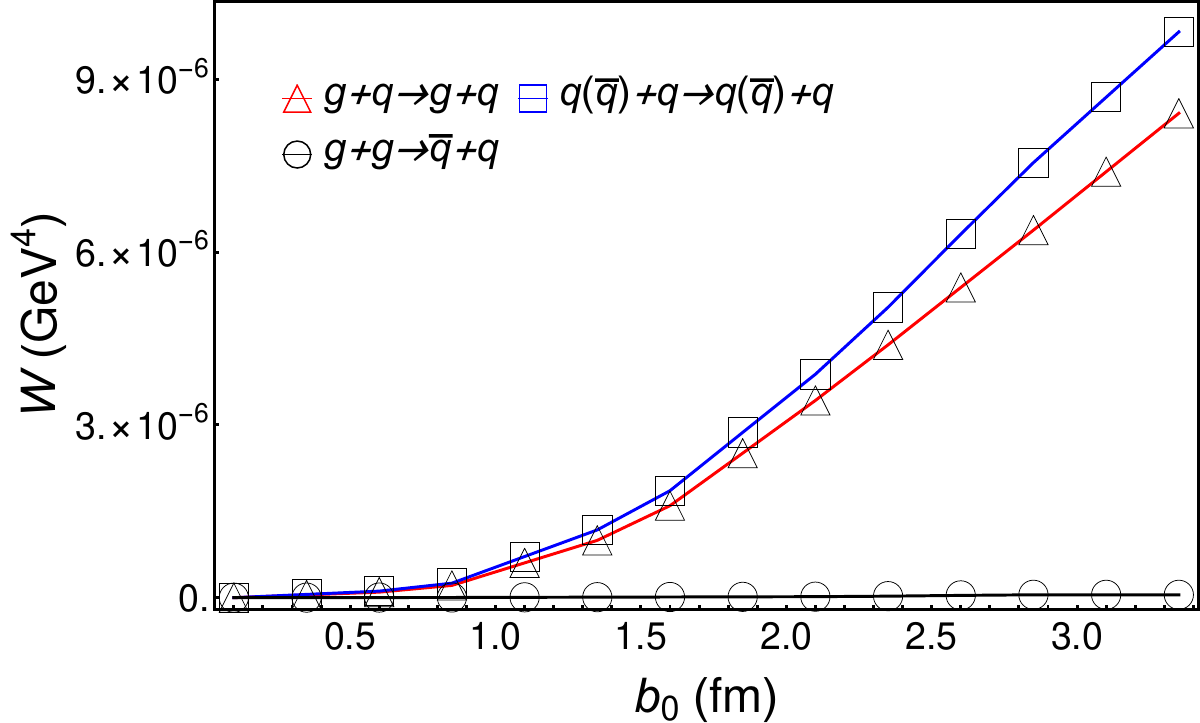}
\par\end{centering}
\caption{Numerical results for $W$ as a function of $b_{0}$, the cutoff for
$|\mathbf{b}_{c}|$. \label{fig:Numerical-results-for}}
\end{figure}

However, it is still an open question whether the QGP reaches
a local equilibrium of the spin degrees of freedom. A microscopic model
of the spin polarization generated by spin--orbit coupling in
particle collisions has been proposed \citep{Zhang:2019xya}.
Without assuming a local equilibrium of spins, it uses an effective
method of wave packets to handle particle scattering for specified
impact parameters. The spin--vorticity coupling naturally emerges from
the spin--orbit coupling encoded in the polarized scattering amplitudes of collisional
integrals when a local equilibrium of the momentum is assumed. First, the collision rate is calculated: 
\begin{eqnarray}
 R_{AB\rightarrow12} = n_{A}n_{B}|v_{A}-v_{B}|\sigma = \frac{d^{3}p_{A}}{(2\pi)^{3}}\frac{d^{3}p_{B}}{(2\pi)^{3}}  f_{A}(x_{A},p_{A})f_{B}(x_{B},p_{B})|v_{A}-v_{B}|\Delta\sigma,
 \label{eq:collision-rate}
\end{eqnarray}
where $v_{A}=|\mathbf{p}_{A}|/E_{A}$ and $v_{B}=-|\mathbf{p}_{B}|/E_{B}$
are the longitudinal velocities, with $\mathbf{p}_{A}=-\mathbf{p}_{B}$
in the center-of-mass frame of colliding particles; $f_{A}$ and $f_{B}$
are the phase space distributions of the incident particles $A$ and $B$,
respectively; and $\Delta\sigma$ denotes the infinitesimal element
of the cross section. After incorporating the components of the wave packets,
the matrix elements of 2-to-2 scattering, the spin projection, and the proper
Lorentz transformations, one obtains the polarization production rate
per unit volume: 
\begin{widetext}
\begin{eqnarray}
\frac{d^{4}\mathbf{P}_{AB\rightarrow12}(X)}{dX^{4}} & = & \frac{1}{(2\pi)^{4}}\int\frac{d^{3}p_{c,A}}{(2\pi)^{3}2E_{c,A}}\frac{d^{3}p_{c,B}}{(2\pi)^{3}2E_{c,B}}\frac{d^{3}p_{c,1}}{(2\pi)^{3}2E_{c,1}}\frac{d^{3}p_{c,2}}{(2\pi)^{3}2E_{c,2}}\nonumber \\
 &  & \times|v_{c,A}-v_{c,B}|G_{1}G_{2}\int d^{3}k_{c,A}d^{3}k_{c,B}d^{3}k_{c,A}^{\prime}d^{3}k_{c,B}^{\prime}\nonumber \\
 &  & \times\phi_{A}(\mathbf{k}_{c,A}-\mathbf{p}_{c,A})\phi_{B}(\mathbf{k}_{c,B}-\mathbf{p}_{c,B})\phi_{A}^{*}(\mathbf{k}_{c,A}^{\prime}-\mathbf{p}_{c,A})\phi_{B}^{*}(\mathbf{k}_{c,B}^{\prime}-\mathbf{p}_{c,B})\nonumber \\
 &  & \times\delta^{(4)}(k_{c,A}^{\prime}+k_{c,B}^{\prime}-p_{c,1}-p_{c,2})\delta^{(4)}(k_{c,A}+k_{c,B}-p_{c,1}-p_{c,2})\nonumber \\
 &  & \times\int d^{2}\mathbf{b}_{c}f_{A}\left(X_{c}+\frac{y_{c,T}}{2},p_{c,A}\right)f_{B}\left(X_{c}-\frac{y_{c,T}}{2},p_{c,B}\right)\nonumber \\
 &  & \times\exp\left[i(\mathbf{k}_{c,A}^{\prime}-\mathbf{k}_{c,A})\cdot\mathbf{b}_{c}\right]\nonumber \\
 &  & \times\sum_{s_{A},s_{B},s_{1},s_{2}}2s_{2}\mathbf{n}_{c}\mathcal{M}\left(\{s_{A},k_{c,A};s_{B},k_{c,B}\}\rightarrow\{s_{1},p_{c,1};s_{2},p_{c,2}\}\right)\nonumber \\
 &  & \times\mathcal{M}^{*}\left(\{s_{A},k_{c,A}^{\prime};s_{B},k_{c,B}^{\prime}\}\rightarrow\{s_{1},p_{c,1};s_{2},p_{c,2}\}\right),\label{eq:rate-pol}
\end{eqnarray}
\end{widetext}
where $\mathbf{P}_{AB\rightarrow12}$ denotes the polarization vector
of particle 2; $\phi_{A}$ and $\phi_{B}$ are the wave packets
for $A$ and $B$, respectively; $\mathbf{n}_{c}=\hat{\mathbf{b}}_{c}\times\hat{\mathbf{p}}_{c,A}$
is the direction of the reaction plane in the center-of-mass frame,
with $\hat{\mathbf{b}}_{c}$ being the unit impact parameter vector;
$f_{A}$ and $f_{B}$ are the distributions at the coordinates $X_{c}+y_{c,T}/2$
and $X_{c}-y_{c,T}/2$, respectively; and $\mathcal{M}$ denotes the amplitude
of the 2-to-2 scattering with all the spins and momenta specified. All
the momenta are defined in the center-of-mass frame and indicated by the index
c. For more details on (\ref{eq:rate-pol}), see Ref. \citep{Zhang:2019xya}.
The wave packets ensure that the colliding particles have a non-vanishing initial angular momentum; the matrix elements encode the collision probability,
and the spin projection and Lorentz transformation provide a consistent
treatment of particle scattering in the thermal bath frame. One can apply
(\ref{eq:rate-pol}) to the quark--gluon system. Then the quark
polarization rate per unit volume with all the 2-to-2 parton scatterings
in a locally thermalized QGP in momentum is 
\begin{eqnarray}
\frac{d^{4}\mathbf{P}_{q}(X)}{dX^{4}} & = & \frac{\pi}{(2\pi)^{4}}\frac{\partial(\beta u_{\rho})}{\partial X^{\nu}}\sum_{A,B,1}\int\frac{d^{3}p_{A}}{(2\pi)^{3}2E_{A}}\frac{d^{3}p_{B}}{(2\pi)^{3}2E_{B}}\nonumber \\
 &  & \times|v_{c,A}-v_{c,B}|[\Lambda^{-1}]_{\;j}^{\nu}\mathbf{e}_{c,i}\epsilon_{ikh}\hat{\mathbf{p}}_{c,A}^{h}\nonumber \\
 &  & \times f_{A}\left(X,p_{A}\right)f_{B}\left(X,p_{B}\right)\left(p_{A}^{\rho}-p_{B}^{\rho}\right)\Theta_{jk}(\mathbf{p}_{c,A})\nonumber \\
 & \equiv & \frac{\partial(\beta u_{\rho})}{\partial X^{\nu}}\mathbf{W}^{\rho\nu},\label{eq:diff-rate}
\end{eqnarray}
which is proportional to the thermal vorticity. Here the tensor $\mathbf{W}^{\rho\nu}$
contains 64 components, each of which involves a 16-dimensional integration.
The numerical calculation of $\mathbf{W}^{\rho\nu}$ is challenging
owing to the very large number of scattering amplitudes and high-dimensional
collision integrals. To tackle this problem,
a new Monte-Carlo integration algorithm, ZMCintegral, which can handle 16-dimensional integration, has been developed for use on multiple GPUs
\citep{Wu_2020cpc_wzpw,Zhang_2020cpc_zw}. The most recent application
of this algorithm is the solution of the Boltzmann equations of a partonic
system \citep{Zhang:2019uor}. The numerical result shows that $\mathbf{W}^{\rho\nu}$
has an antisymmetric form: 
\begin{eqnarray}
\mathbf{W}^{\rho\nu} & = & \left(\begin{array}{cccc}
0 & 0 & 0 & 0\\
0 & 0 & W\mathbf{e}_{z} & -W\mathbf{e}_{y}\\
0 & -W\mathbf{e}_{z} & 0 & W\mathbf{e}_{x}\\
0 & W\mathbf{e}_{y} & -W\mathbf{e}_{x} & 0
\end{array}\right),
\end{eqnarray}
where $W$ is approximately constant. Note that $W$
depends on the cutoff of the impact parameter, as shown in Fig. \ref{fig:Numerical-results-for}.
Finally, the polarization rate per unit volume for one quark flavor
can be expressed compactly: 
\begin{equation}
\frac{d^{4}\boldsymbol{P}_{q}(X)}{dX^{4}}=2W\nabla_{X}\times\left(\frac{\boldsymbol{u}}{T}\right).
\end{equation}
This is a good example of how spin--vorticity coupling emerges
naturally from particle scattering.

Note that the result of Ref. \citep{Zhang:2019xya} or Eq.
(\ref{eq:diff-rate}) does not include the back reaction to contain
the growing polarization with increasing vorticity in the absence of a cutoff.
A systematic derivation of the spin polarization from the vorticity through
non-local collisions with back reactions has been performed
\citep{Weickgenannt:2020aaf} by expanding the collision terms
in the Planck constant for massive fermions \citep{Yang:2020hri}.

\subsection{Experimental results \label{subsec:Recent-experimental-results}}

\begin{figure}[!tb]
\includegraphics[scale=0.40]{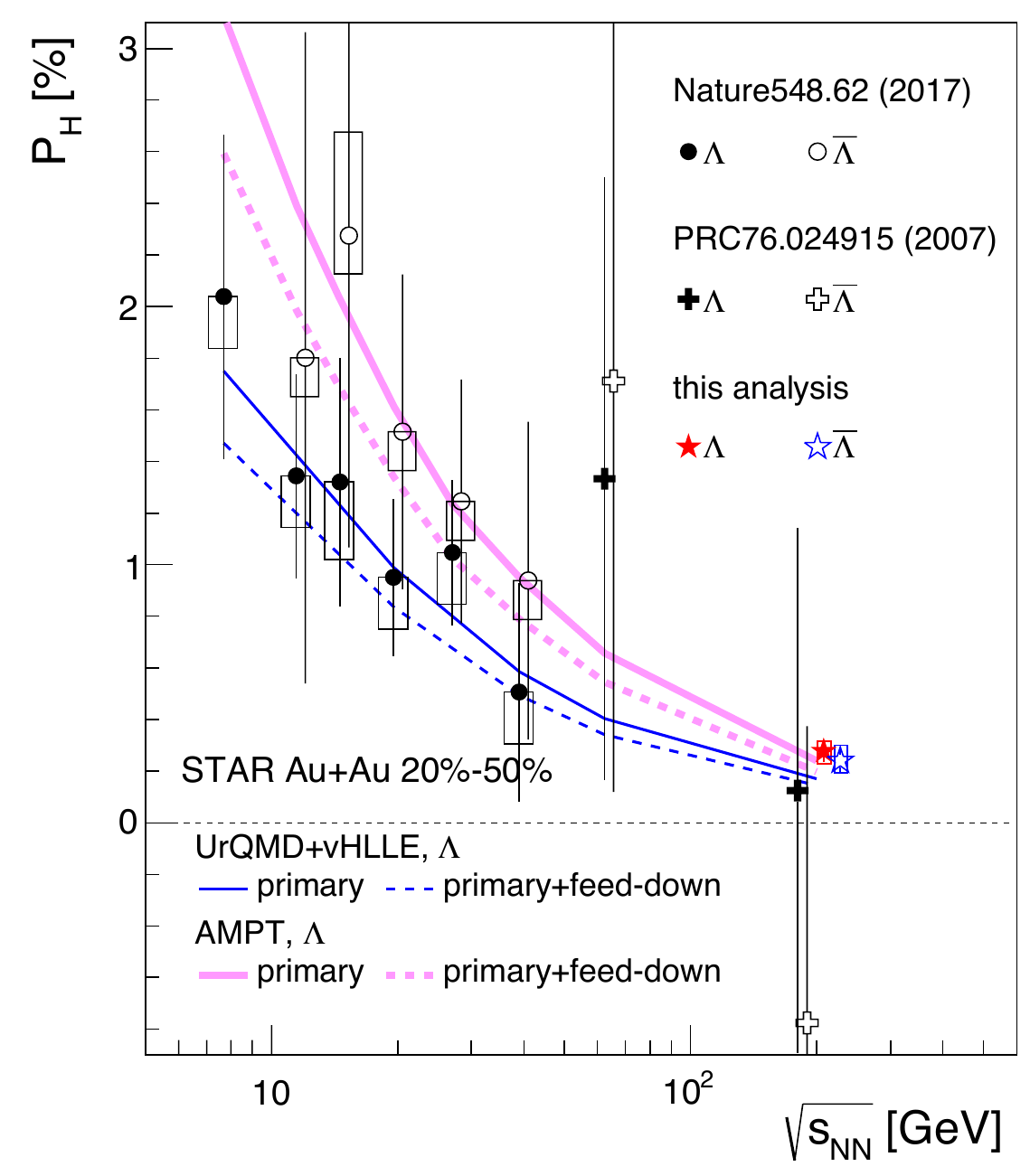}
\includegraphics[scale=0.5]{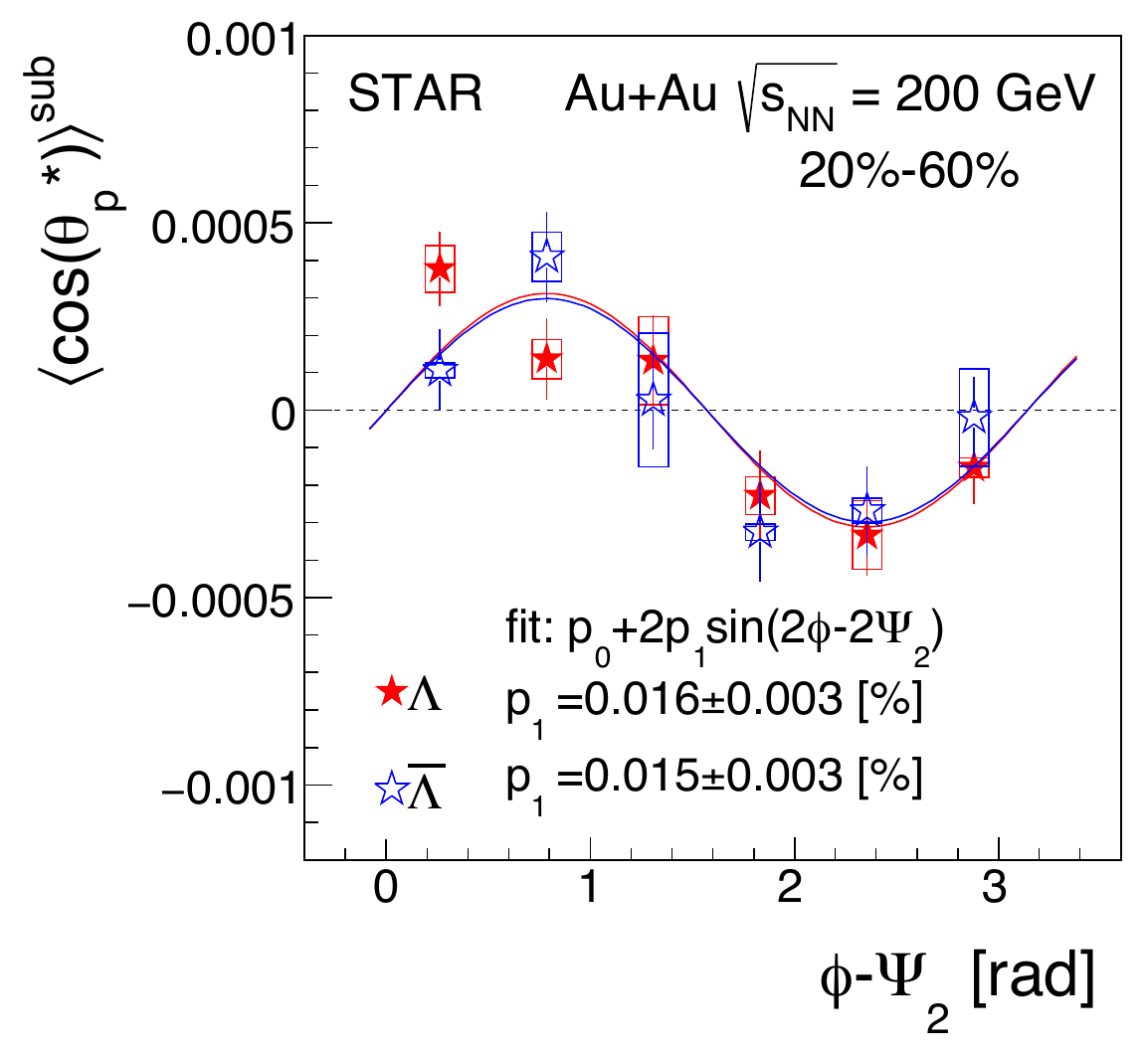}
\caption{Global polarization of $\Lambda$ and $\bar{\Lambda}$ (left panel)
\citep{STAR:2017ckg,Adam:2018ivw} and local polarization as a function
of azimuthal angle $\phi$ relative to the second-order event plane
$\Psi_{2}$ (right panel) \citep{Adam:2019srw} measured by STAR.
\label{fig:Global-polarization}}

\end{figure}

The global polarization of $\Lambda$ hyperons can be measured through
their weak decays. The angular distribution of the daughter baryons is
\citep{STAR:2017ckg,Adam:2018ivw} 
\begin{equation}
\frac{dN}{d\cos\theta^{*}}=\frac{1}{2}\left(1+\alpha_{H}P_{H}\cos\theta^{*}\right),
\end{equation}
where $\alpha_{H}$ is the decay constant of the hyperon, $P_{H}$
is the polarization of the hyperon (a fraction), and $\theta^{*}$ is the
angle between the momentum of the daughter baryon and the polarization
direction in the hyperon rest frame. The experimental results for
the global polarization of $\Lambda$ and $\bar{\Lambda}$ hyperons
measured by the STAR Collaboration \citep{STAR:2017ckg,Adam:2018ivw}
are shown in the left panel of Fig. \ref{fig:Global-polarization}.
$P_{H}(\Lambda)$ and $P_{H}(\bar{\Lambda})$ decrease
with the collision energy. From the data, the fluid
vorticity can be estimated as $\omega\simeq k_{B}T(P_{\Lambda}+P_{\bar{\Lambda}})/\hbar$
\citep{Becattini_PRC2017_bkimuv}. It also appears that $P_{H}(\bar{\Lambda})\gtrsim P_{H}(\Lambda)$,
although the difference is not significant; this result may suggest a possible contribution from
the magnetic field or other effects \citep{Csernai:2018yok,Guo:2019mgh}.
The data can be described by various
phenomenological models using (\ref{eq:PL_vector}) \citep{Karpenko_epjc2017_kb,XieYilong_prc2017_xwc,LiHui_prc2017_lpwx,Sun:2017xhx,WeiDexian_prc2019_wdh,Shi_plb2019_sll}.

The right panel of Fig. \ref{fig:Global-polarization} shows the recent
STAR Collaboration measurement \citep{Adam:2019srw} of the longitudinal polarization
as a function of the azimuthal angle $\phi$ relative to the second-order
event plane $\Psi_{2}$. The longitudinal spin polarization
data show positive $\sin(2\phi-2\Psi_{2})$ behavior, whereas the theoretical
results of the relativistic hydrodynamics model \citep{Becattini_prl2018_bk}
and transport models \citep{XieYilong_prc2017_xwc,Xia:2018tes,WeiDexian_prc2019_wdh}
show the opposite sign along the beam line direction. A simulation
using CKT \citep{Liu:2019krs} and the results obtained using a simple phenomenological
model \citep{Voloshin_2018epjWeb} give the correct sign
as the data. The azimuthal angle dependence of the spin polarization
in the direction of the global OAM has been measured by the STAR Collaboration.
Some phenomenological models do not well describe the STAR data for
the azimuthal angle dependence of the global polarization.

The sign problem of the local polarization requires further investigations.
It may indicate a need for new frameworks to describe the spin dynamics,
such as the quantum spin kinetic theory for massive particles (see
Sec. \ref{sec:Quantum-kinetic-theory}) or the spin hydrodynamics
(see Sec. \ref{subsec:Several-theoretical-updates}). 

One possible effect is that of feed-down decays. The hyperons measured
in experiments may be produced by decays of heavier resonance particles.
However, the authors of Refs. \citep{Xia:2019fjf,Becattini_epjc2019_bcs} concluded that feed-down effects decreased by approximately $10\%$
for the $\Lambda$ primordial spin polarization; thus, they do not solve
the spin sign problem. 

In addition, the sign problem of the local polarization may also
indicate that the assumption of global or local equilibrium of the spin may
not be justified, and thus the thermal vorticity may not be the correct quantity
for the spin chemical potential. Wu et al. \citep{Wu:2019eyi, Wu:2020yiz}
tested four different types of vorticity: the kinematic
vorticity, relativistic extension of the non-relativistic vorticity,
temperature vorticity, and thermal vorticity. They calculated
the local polarization of hyperons corresponding to each type of vorticity.
By using the $(3+1)$-dimensional hydrodynamic model with the AMPT initial
conditions encoding the global OAM, they found a few remarkable differences between the results for different vorticities. First, although all four types of vorticity
give the correct sign and magnitude of the polarization along the global
OAM direction, only the temperature vorticity agrees with the STAR preliminary result in showing a decreasing
trend of the azimuthal angle dependence.
Second, only the temperature vorticity can simultaneously provide the correct sign
and magnitude of the longitudinal polarization. This result
suggests the possibility of spin--temperature vorticity coupling,
analogous to magnetic moment--magnetic
field coupling. It is also possible that the agreement may be a coincidental
result of the main assumption that the spin vector depends on the temperature
vorticity in the same way as it does on the thermal vorticity. Further investigation
is needed to clarify why the results for the temperature vorticity are the most satisfactory.

The vector meson spin alignment is another recent research topic.
The spin alignment of a vector meson is described by the $3\times3$
Hermitian spin-density matrix \citep{Schilling:1969um,Abelev:2008ag}.
The 00-component of the spin-density matrix enters the angular distribution
of its decay daughter as 
\begin{equation}
\frac{dN}{d\cos\theta^{\text{\textasteriskcentered}}}\propto[(1-\rho_{00})+(3\rho_{00}-1)\cos^{2}\theta^{\text{\textasteriskcentered}}],\label{eq:vector_01}
\end{equation}
where $\theta^{\ast}$ is the angle between the decay daughter and
the spin quantization direction in the vector meson's rest frame.
Thus, the $\rho_{00}$ value of the vector meson can be measured using the angular
distribution of its decay daughter. The vector
meson ($K^{\ast0}$ and $\phi$) spin alignments have been measured by
the ALICE Collaboration \citep{Acharya:2019vpe}. The $\rho_{00}$
value is consistent with $1/3$ for both $K^{*0}$ and $\phi$ mesons in
p+p collisions. In Pb+Pb collisions, the $\rho_{00}$ value of $K^{*0}$
is approximately $1/3$ at high $p_{T}$ and less than $1/3$ at low $p_{T}$.

Theoretical calculations using the statistical-hydro model \citep{Becattini_PRC2017_bkimuv}
and quark coalescence model \citep{Yang:2017sdk} give 
\begin{equation}
\rho_{00}^{\phi}\approx\frac{1}{3}-\frac{4}{9}P_{\Lambda}P_{\bar{\Lambda}}\lesssim\frac{1}{3}.
\end{equation}
Using the STAR data, $P_{\Lambda}\approx(1.08\pm0.15\pm0.11)\%$ and
$P_{\bar{\Lambda}}\approx(1.38\pm0.30\pm0.13)\%$ \citep{STAR:2017ckg,Adam:2018ivw},
one can estimate $(4/9)P_{\Lambda}P_{\bar{\Lambda}}\approx6.6\times10^{-5}$,
which may suggest that $\rho_{00}^{\phi}$ cannot be significantly
larger than 1/3 and thus contradict the STAR preliminary data for $\rho_{00}^{\phi}$. 

\begin{figure}[tb]
\includegraphics[scale=0.35]{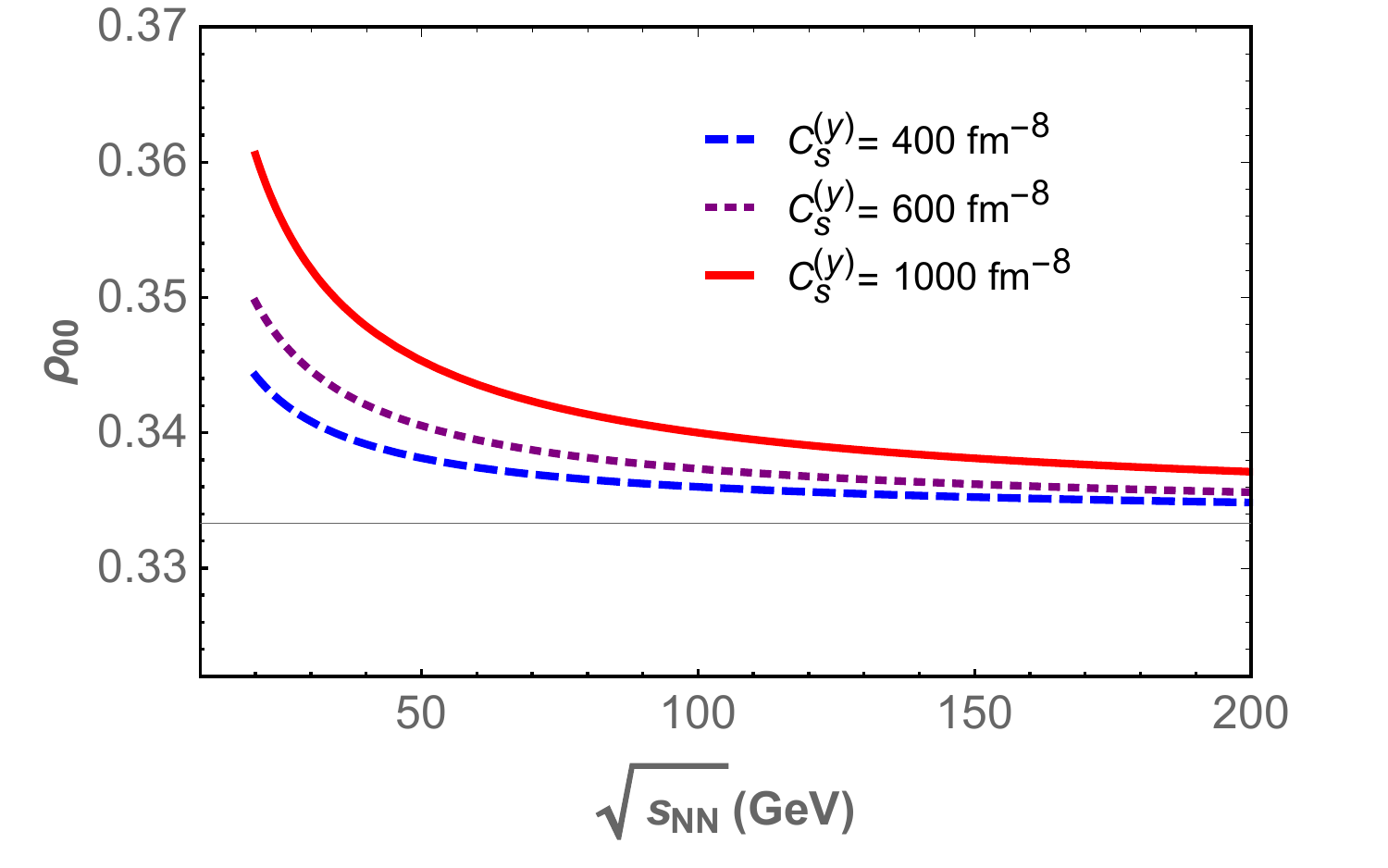}
\caption{Theoretical prediction of $\rho_{00}^{\phi}$ in heavy-ion collisions
as a function of collision energy. The thin horizontal solid line
shows the no-alignment value, $\rho_{00}=1/3$. Three values of $C_{s}^{(y)}$
are chosen \citep{Sheng:2019kmk}. \label{fig:rho-00-figure}}
\end{figure}

To solve this puzzle, it is proposed that a strangeness current can
exist in heavy-ion collisions and give rise to a non-vanishing mean
$\phi$ field \citep{Sheng:2019kmk}. Like the magnetic field, the
magnetic part of the $\phi$ field can also polarize $s$ and $\bar{s}$
through their magnetic moments; this polarization contributes to the polarization
of $\Lambda$ and $\bar{\Lambda}$, whereas the contribution from the
electric part of the $\phi$ field is absent and therefore is not
constrained by the polarization of $\Lambda$ and $\bar{\Lambda}$.
However, the electric part of the $\phi$ field contributes significantly
to $\rho_{00}^{\phi}$, which is positive definite \citep{Sheng:2019kmk}:
\begin{equation}
\rho_{00}^{\phi}\approx\frac{1}{3}+\frac{g_{\phi}^{2}}{27m_{s}^{4}T_{\mathrm{eff}}^{2}}\left\langle \mathbf{p}^{2}\right\rangle _{\phi}\left\langle E_{\phi,z}^{2}+E_{\phi,x}^{2}\right\rangle ,
\end{equation}
where $\mathbf{E}_{\phi}$ is the electric part of the $\phi$ field,
$m_{s}$ is the strange quark mass, $\left\langle \mathbf{p}^{2}\right\rangle _{\phi}$
is the mean value of $\mathbf{p}^{2}$ for $s$ or $\bar{s}$ in the
$\phi$ meson wave function, $g_{\phi}$ is the coupling constant
of the $s$ quark to the $\phi$ meson in the quark--meson model, and
$T_{\mathrm{eff}}$ is the average temperature of the fireball. The
contribution originates from the spin--orbit coupling term that polarizes
$s$ and $\bar{s}$. Thus, one can see that $\rho_{00}$ for the $\phi$
meson is a good tool for analyzing the $\phi$ mean field even if it may
strongly fluctuate in space-time \citep{Sheng:2019kmk}. The theoretical
prediction of $\rho_{00}^{\phi}$ as a function of the collision energy
is shown in Fig. \ref{fig:rho-00-figure}.

However, this theory does not work for another vector meson, $K^{*0}$,
for several reasons. First, because of the unequal masses of $\bar{s}$
and $d$, one cannot derive the same formula as that for $\phi$ mesons, in
which the contributions from the vorticity and those from the electric and magnetic
fields are decoupled. Second, the interaction of $K^{*0}$ with the
surrounding matter is much stronger than that of the $\phi$ meson. The above
points are supported by preliminary data from ALICE \citep{Acharya:2019vpe}.

\section{Summary \label{sec:Summary}}

We gave a brief overview of recent theoretical developments
and experimental results on the effects of chirality and vorticity
on heavy-ion collisions. We focus on works reported at
the Quark Matter conference of 2019. 

We discussed the time evolution of magnetic fields in various models of the QGP, for example, the Lienard--Wiechert potential
and MHD models. Macroscopic models such as the second-order
dissipative MHD and AVFD models are applied in phenomenological
studies. 

It is still experimentally challenging to obtain a clear CME signal
against the dominant backgrounds. The non-flow correlations for the CMW give
rise to additional backgrounds in the slope of $\Delta v_{2}(A_{\textrm{ch}})$.
More detailed studies of the CME and CMW are needed. 

The kinetic theory of massive fermions has been developed for the covariant
and equal-time Wigner functions and the Schwinger--Keldysh formalism.
The collision terms have been studied at the leading and next-to-leading
order in an expansion of the Planck constant. 

The decomposition of the total angular momentum into the OAM and spin
has been debated for quite a long time. Another important issue that remains to be resolved is the relationship between the theoretical 
spin and the results of experimental measurements. 

For the global polarization of $\Lambda$ and $\bar{\Lambda}$ hyperons,
various phenomenological models give consistent descriptions of the experimental
data. For the local polarization, the nature of the spin sign problem
is still unclear. It has been proposed that a significant positive
deviation of $1/3$ for $\rho_{00}^{\phi}$ may indicate the existence
of a mean $\phi$ field. 

\section*{ACKNOWLEDGMENTS}

J.-H.G. is supported in part by the National Natural Science Foundation
of China under Grant No. 11890713. G.-L.M. is supported by the National
Natural Science Foundation of China under Grant Nos. 11890714, 
11835002, 11961131011, and 11421505, and the Strategic Priority
Research Program of the Chinese Academy of Sciences under Grant No. XDB34030202.
Q.-W. is supported in part by the National Natural Science Foundation
of China under Grant Nos. 11535012 and 11890713, and the
Strategic Priority Research Program of the Chinese Academy of Sciences
under Grant No. XDB34030102. 


\end{document}